\documentclass[12pt, draftclsnofoot, onecolumn]{IEEEtran}
\usepackage{graphicx}
\usepackage{caption}
\usepackage[ruled]{algorithm2e}
\ifCLASSOPTIONcompsoc
  \usepackage[caption=false,font=normalsize,labelfont=sf,textfont=sf]{subfig}
\else
  \usepackage[caption=false,font=footnotesize]{subfig}
\fi
\usepackage{indentfirst}

\usepackage{amsmath}
\allowdisplaybreaks[4]
\usepackage{amssymb}
\usepackage{times}
\usepackage{mathtools}
\usepackage{psfrag}
\usepackage{cite}
\usepackage{lastpage}
\usepackage{fancyhdr}
\usepackage{color}
 \usepackage{amsthm}
\usepackage{bigints}
\newtheorem{Theorem}{Theorem}
\newtheorem{Definition}{Definition}

\newtheorem{Proposition}{Proposition}

\newtheorem{Lemma}{Lemma}

\newtheorem{Corollary}{Corollary}
\newtheorem{theorem}[Theorem]{$\mathbf{Theorem}$}

\newcommand{\Equ}[1]{
  \begin{align}
    #1
  \end{align}}

\newcommand{\SubEquL}[2]{
  \begin{subequations}\label{#1}
    \begin{align}
     #2
    \end{align}
\end{subequations}}

\begin{document}
\title{On the Application of Quasi-Degradation to Network NOMA in Downlink CoMP Systems}
\author{Yanshi Sun, Zhiguo Ding, \IEEEmembership{Fellow, IEEE}, Xuchu Dai, Momiao Zhou,\\ Zhizhong Ding
\thanks{
Y. Sun, M. Zhou and Z. Ding are  with the School of Computer Science and Information
Engineering, Hefei University of Technology, Hefei, 230009, China. (email: \{sys, mmzhou, zzding\}@hfut.edu.cn).
 
Z. Ding is with the School of Electrical and Electronic Engineering, the University of Manchester, Manchester M13 9PL, U.K. (email:zhiguo.ding@manchester.ac.uk).

X. Dai is with the CAS Key Laboratory of Wireless-Optical Communications, University of Science and Technology of China, Hefei, 230026, China. (email: daixc@ustc.edu.cn).
}\vspace{-2em}}
\maketitle

\begin{abstract}
The application of network non-orthogonal multiple access (N-NOMA) technique to coordinated multi-point (CoMP) systems has attracted significant attention due to its superior capability to improve connectivity and maintain reliable transmission for CoMP users simultaneously.
Based on the concept of quasi-degraded channel for N-NOMA, this paper studies the precoding design for downlink N-NOMA scenarios with two base stations (BSs) equipped with multiple antennas.
In specific, under quasi-degraded channels, simple linear precoding based N-NOMA can achieve the same
minimal total transmission power as theoretically optimal but complicated dirty paper coding (DPC) scheme, when the users' target rates and minimal transmission power of each BS are given.
In this paper, the channel quasi-degradation (QD) condition is first rigorously derived for the scenario with single CoMP user and two NOMA users. The closed-form optimal precoders for N-NOMA under quasi-degraded channels are also provided. Then, based on QD condition, a novel hybrid N-NOMA (H-N-NOMA) scheme is proposed, which is a mixture of N-NOMA and conventional zero-forcing beamforming (ZFBF) scheme.
Further, for the scenarios with more users, a low-complexity QD based user pairing (QDUP) algorithm is proposed. Numerical results are presented to reveal the impact factors of QD channels, and also demonstrate the superior performance of the proposed H-N-NOMA/QDUP scheme. It is shown that the proposed H-N-NOMA/QDUP scheme can effectively exploit the benefit of multi user diversity.
\end{abstract}
\begin{IEEEkeywords}
Network non-orthogonal multiple access (N-NOMA), coordinated multi-point (CoMP), quasi-degradation, precoding, minimal transmission power
\end{IEEEkeywords}
\section{Introduction}
Coordinated multi-point (CoMP) techniques, which utilize the coordination among multiple spatially distributed base stations (BSs), play an important role in ensuring the transmission reliability of  cell-edge users\cite{irmer2011coordinated,kim2021energy}. However, conventional CoMP techniques are designed based on orthogonal multiple access (OMA), where different users are allocated with orthogonal channel resource blocks, resulting in low spectral efficiency. For example, when multiple BSs cooperatively serve a cell-edge user via CoMP, each BS needs to allocate a resource block for this user, and the resource block cannot be accessed by other users. Thus, the spectral resource becomes more stringent as the number of cell-edge users increases.

Recently, non-orthogonal multiple access (NOMA) has raised tremendous attention from both academia and industry, due to its superior spectral efficiency and support for massive connectivity\cite{ding2017NOMAsurvey}. The key idea of NOMA is to encourage multiple users share the same resource block. Network NOMA (N-NOMA) constitutes an important branch of NOMA, which can significantly ease the aforementioned dilemma caused by OMA based CoMP\cite{choi2014non, sys2017nnomafeasibility,new2020network,ali2018coordinated}. In N-NOMA, multiple BSs cooperatively serve cell-edge users (termed CoMP users) via conventional CoMP techniques, meanwhile, each BS serves additional cell-center users (termed NOMA users) by using the same resource blocks allocated to the CoMP users. Compared with OMA based CoMP schemes, N-NOMA
can support larger connectivity and improve spectral efficiency, while the CoMP users' performance can be guaranteed. Due to this appealing advantage, N-NOMA has been recognized as an valuable research topic for
5G and beyond\cite{makki2020survey}.
\subsection{Related work}
According to different antenna configurations, N-NOMA can be classified into two types: (a) single input single output (SISO) N-NOMA, where BSs and users are equipped with single antenna;
(b) multiple input multiple output (MIMO) N-NOMA, where BSs and (or) users are equipped with multiple antennas. In the following, related works on SISO and MIMO N-NOMA are described respectively.
\subsubsection{SISO N-NOMA}
On the one hand, some papers focused on statistic performance analysis of N-NOMA. In \cite{choi2014non}, the ergodic rate achieved by Alamouti coding based N-NOMA scheme was studied. In \cite{tian2016performance}, the outage performance of a opportunistic N-NOMA scheme with user cooperation
was investigated. In \cite{sys2017nnomafeasibility}, distributed analog beamforming was applied in a three-BS N-NOMA scenario. In \cite{al2019generalized}, a generalized N-NOMA scheme was proposed, where the cell-center users can also be cooperatively served. The concept of mutual successive interference cancellation (SIC) was proposed for further performance improvement of N-NOMA. In \cite{sys2019PCP,zhang2020performance,elhattab2020comp,sun2021outage}, stochastic geometry was applied to model and analyze the performance of N-NOMA by considering the interference from the whole network.
On the other hand, some papers focused on resource optimization for N-NOMA under different kinds of constraints. In \cite{liu2017power,ali2018downlink,elhattab2020power}, power allocation algorithms were proposed for N-NOMA. In \cite{rezvani2021resource}, a joint power allocation, user pairing and coordinated scheduling method was studied for a multi-connectivity based virtual N-NOMA system. In \cite{elhattab2022joint}, the joint user clustering and power allocation issue was investigated.
Besides, the combination of N-NOMA with advanced wireless techniques were also investigated \cite{hedayati2020comp,lei2020outage,elhattab2020joint,elhattab2022ris,wang2020power,hou2021joint}.
\subsubsection{MIMO N-NOMA}
When the cooperating BSs are equipped with multiple antennas, by applying proper precoding (or beamforming) design, the spatial degree of freedom can be adequately utilized and the inter-cell interference (ICI) can be effectively mitigated. In \cite{shin2016coordinated}, two interference alignment based downlink N-NOMA precoding methods were proposed for the case with full channel state information (CSI) and partial CSI, respectively. \cite{nguyen2017precoder} proposed an optimal precoding algorithm to maximize the sum throughput of a downlink N-NOMA system with two cooperating BSs. In \cite{sun2018joint}, the precoding design was considered for the scenario with multiple cooperating BSs. The users served by each BS are divided into two groups, and the proposed precoding algorithm aims to maximum the rates achieved by one group while guarantee the performance of the users belonging to the other group. For a similar scenario as in \cite{sun2018joint}, in \cite{fu2020zero}, an efficient algorithm was proposed to minimize the total transmission power.
\subsection{Motivation and contribution}
The drawbacks of existing work on precoding design for MIMO N-NOMA are as follows.
\begin{itemize}
\item Existing work only provides statistical comparison between N-NOMA and traditional schemes. However, in reality, under some channel conditions, N-NOMA might not perform better than traditional schemes, such as zero-forcing beamforming (ZFBF). Thus, dynamically determining whether adopting N-NOMA as the transmission strategy according to channel conditions can further improve the spectral efficiency. Unfortunately, the characteristic of the channels which are favorable for the application of N-NOMA are still veiled, making it difficult to find a simple but effective rule for switching between N-NOMA and other schemes.
\item Most existing precoding methods for N-NOMA are relied on iterative algorithms, which heavily lacks of efficiency for practical application, as well as insights for the better understanding of N-NOMA.
\item Most of the papers focus on non-joint transmission (NJT), where the CoMP user's signal is transmitted by only one BS and the intension of coordination among BSs is just to mitigate ICI. However, NJT cannot cope well with the scenarios where cell-edge users face severely weak channels or their demand for quality of service (QoS) are pretty high. Thus, taking joint transmission (JT) into consideration for precoding design of N-NOMA is an urgent and necessary task.
\end{itemize}

To fulfill the above issues, this paper intends to study the relationship between N-NOMA and optimal dirty paper coding (DPC) scheme, by applying and extending the concept of quasi-degradation (QD) which is originated from the research for single-cell MISO NOMA \cite{chen2016application,zhu2020optimal,liu2021quasi}. To be specific, when the channels satisfy QD condition, simple linear precoding based N-NOMA can achieve the same
minimal total transmission power as theoretically optimal but complicated DPC scheme, when the users' target rates and minimal transmission power of each BS are given. Thus, studying quasi-degraded channels can help to reveal the properties of the channels which are favorable for the application of N-NOMA. In this paper, QD condition is explicitly given for downlink N-NOMA with two BSs, based on which efficient N-NOMA schemes are then proposed. The main contributions of this paper are listed as follows.
\begin{itemize}
\item The optimal precoding design problem for N-NOMA with two BSs is formulated, which aims to minimize the total transmission power under the constraints on users' target rates and the maximum transmission power of each BS. Different from existing N-NOMA precoding methods which only focus on NJT, this paper deals with a more general case by jointly considering JT and NJT.
\item The comparison between the optimal design for N-NOMA and DPC is studied. Particularly, the definition of QD for N-NOMA is formally defined. Note that the concept of QD for N-NOMA is different from that of single cell NOMA\cite{chen2016application}. Because in single cell NOMA as in \cite{chen2016application}, power constraint of a single BS is not considered, while in N-NOMA, power constraint of a single BS must be considered to avoid the consequence that the CoMP user is always served by one BS. Thus, it is more difficult to obtain the QD condition of N-NOMA than that of single cell NOMA. Fortunately, through rigorous derivation,  closed-form expression for QD condition of N-NOMA is provided. Besides, closed-form optimal precoder of N-NOMA under quasi-degraded channel is also obtained.
\item  Based on the obtained QD condition, an novel hybrid N-NOMA (H-N-NOMA) scheme is proposed, which can dynamically choose N-NOMA and ZFBF as the transmission strategy according to real-time CSI. By taking the advantage of closed-form expressions, the proposed H-N-NOMA scheme can significantly reduce the complexity compared to existing iterative algorithms, and the performance can be guaranteed.
\item An efficient QD based user pairing (QDUP) algorithm for scenarios with multiple CoMP users is proposed.
In QDUP, users whose channels satisfying QD condition are preferentially grouped. The complexity of the proposed QDUP is much lower than that of exhaustive search.
Combining H-N-NOMA and QDUP, an novel transmission scheme called H-N-NOMA/SUPA is then proposed.
Numerical results shows that the proposed  H-N-NOMA/SUPA scheme can significantly outperform existing schemes. It is also shown that  the proposed  H-N-NOMA/SUPA scheme can effectively exploit the benefit of multi user diversity.
\end{itemize}

The remainder of this paper is organized as follows. Section II gives the system model, which formulates the optimization problem for N-NOMA and DPC. In section III,
the optimal solution of DPC is provided, and the QD condition is obtained. In section IV, by applying the obtained results in Section III, H-N-NOMA and QDUP algorithms are proposed. Section V illustrates the numerical results and Section VI concludes the paper.
\section{System model}
\begin{figure}[!t]
\vspace{-1em}
\setlength{\abovecaptionskip}{0em}   
\setlength{\belowcaptionskip}{-2em}   
\centering
\includegraphics[width=3in]{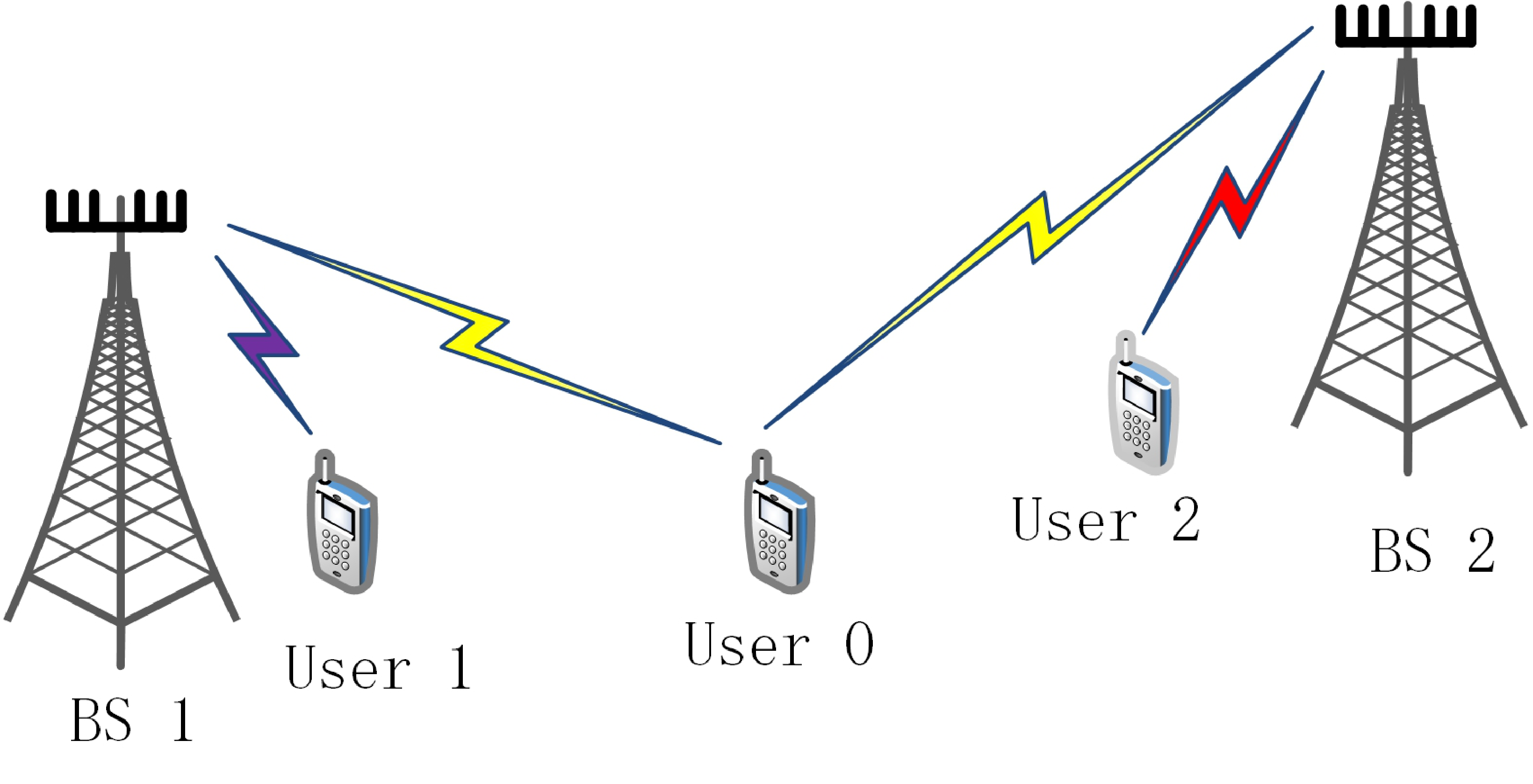}
\caption{Illustration of the system model.}
\label{system_model}
\end{figure}

Consider a two-cell downlink N-NOMA system. As shown in Fig. \ref{system_model}, there are two BSs, which are  termed BS $1$ and BS $2$, respectively.
A CoMP user, termed user $0$,  which is far from both considered BSs, is served cooperatively by the two BSs.
In addition to the CoMP user, each BS individually serves a NOMA user which is close to the BS, by occupying the same resource block allocated to the CoMP user. Particularly, the NOMA user associated with BS $i$ ($i=1,2$) is denoted by user $i$.
Each BS is equipped with $N$ antennas and each user is equipped with a single antenna.  The channel between BS $i$ ($i=1,2$) and user $j$ ($j=0,1,2$) is modeled by $\mathbf{h}_{i,j}$. Note that in this paper, it is assumed that the distance between the two cooperating BSs are far enough so that ICI at the NOMA users can be neglected.

The transmitted signal by BS $i$ ($i=1,2$) is given by:
\begin{align}
x_i=\mathbf{w}_{i0}s_0+\mathbf{w}_{ii}s_i, i=1,2
\end{align}
where $s_i$ ($i=0,1,2$) is the signal intended for user $i$, $\mathbf{w}_{i0}$ ($i=1,2 $) is the beamforming vector for the CoMP user, $\mathbf{w}_{ii}$ ($i=1,2$) is the beamforming vector for user $i$.

Note that, $\mathbf{w}_{i0}$ might be a zero vector. And according to the values of $\mathbf{w}_{i0}$ ($i=1,2$), the transmission scheme can be classified into the following two different types:
\begin{itemize}
  \item JT, where both BSs transmit information to the CoMP user, i.e., $\mathbf{w}_{i0}\neq 0$, $i=1,2$;
  \item NJT, where only one BS transmits information to the CoMP user, i.e., $\exists i, \mathbf{w}_{i0}=0$.
\end{itemize}

The CoMP user treats the signals of NOMA users as interferences, thus the signal to interference plus noise ratio (SINR) of the CoMP user to decode its own signal is:
 \Equ{\label{Ch1_eq_sinr1}
  \mathrm{SINR}_{0\rightarrow0}=\frac{||\mathbf{h}_{10}^{H}\mathbf{w}_{10}||^2+||\mathbf{h}_{20}^{H}\mathbf{w}_{20}||^2}{||\mathbf{h}_{10}^{H}\mathbf{w}_{11}||^2+||\mathbf{h}_{20}^{H}\mathbf{w}_{22}||^2+\sigma ^2}\,}
where $\sigma^2$ is the noise power, $\mathrm{SINR}_{j\rightarrow j'}$ denotes the SINR when user $j$ decodes the signal of user $j'$.

Different from the the CoMP user, if $\mathbf{w}_{i0}\neq 0$, $i=1,2$, the NOMA user (user $1$ or $2$) first decodes the signal of the CoMP user with the following SINR:
\Equ{\label{Ch1_eq_sinr2}
 \mathrm{SINR}_{j\rightarrow0}=\frac{||\mathbf{h}_{ij}^{H}\mathbf{w}_{j0}||^2}{||\mathbf{h}_{ij}^{H}\mathbf{w}_{ij}||^2+\sigma ^2},j=1,2,i=j.}

If the CoMP user's signal can be successfully decoded, user $j$ ($j=1,2$) carries out successive interference cancellation (SIC) to remove the signal of the CoMP user, and then decodes its own signal. If $\mathbf{w}_{i0} = 0$, $i=1,2$, user $i$ will decode its own signal directly. In the above two cases, the SINR when the NOMA user decodes its own signal can be expressed as:
 \Equ{\label{Ch1_eq_sinr3}\mathrm{SINR}_{j\rightarrow j}=\frac{||\mathbf{h}_{ij}^{H}\mathbf{w}_{ij}||^2}{\sigma ^2},j=1,2, i=j.}

Thus, the achievable rates of the users are given by:
 \Equ{\label{Ch1_eq_rate0}
 R_0=\min \bigg\{ \log \left( 1+\mathrm{SINR}_{0\rightarrow0} \right), \underset{{{\underset{j=1,2}{\mathbf{w}_{j0}\ne 0}}}}{\min}\log \left( 1+\mathrm{SINR}_{j\rightarrow0} \right) \bigg\}}
\Equ{\label{Ch1_eq_rate1}R_j=\log \left\{ 1+\mathrm{SINR}_{j\rightarrow j} \right\} ,j=1,2.}
Note that user $0$'s achievable rate depends on not only its own channel condition, but also the NOMA users' channel conditions.

\subsection{Formulation of N-NOMA}
Given the target rate of each user and the largest transmit power of each BS, the total transmission power minimization problem of N-NOMA can be formulated as follows:
\SubEquL{Ch1_P1}{
  \min_{\begin{array}{c}
    \mathbf{w}_{10},\mathbf{w}_{20},
    \mathbf{w}_{11},\mathbf{w}_{22}
  \end{array}} &||\mathbf{w}_{10}||^2+||\mathbf{w}_{20}||^2+||\mathbf{w}_{11}||^2+||\mathbf{w}_{22}||^2\\
           s.t.  \quad\quad & R_j\ge r_j,j=0,1,2
  \\
  &                ||\mathbf{w}_{10}||^2+||\mathbf{w}_{11}||^2\le P_{max},
  \\
  &                 ||\mathbf{w}_{20}||^2+||\mathbf{w}_{22}||^2\le P_{max}.
}
where (\ref{Ch1_P1}b) means that the achievable rate of user $j$ ($j=0,1,2$) must be larger than a given target rate $r_j$, (\ref{Ch1_P1}c) and (\ref{Ch1_P1}d) mean that the transmission power of each BS should not exceed a given maximum $P_{max}$.
\subsection{Formulation of Dirty paper coding (DPC)}
By utilizing the obtained information of each user at transmitters, DPC can effectively avoid inter-user interferences, which generally achieves the optimal performance. Particularly, in the considered DPC in this paper, at BS $1$, user $0$'s information is encoded before user $1$, which ensures that user $1$ can avoid the interference from user $0$. Similarly, user $2$ can also avoid the interference from user $0$. Thus, the power minimization problem of DPC can be given by:
 \SubEquL{Ch1_DPC1}{
  \min_{\begin{array}{c}
    \mathbf{w}_{10},\mathbf{w}_{20},
    \mathbf{w}_{11},\mathbf{w}_{22}
  \end{array}} &||\mathbf{w}_{10}||^2+||\mathbf{w}_{20}||^2+||\mathbf{w}_{11}||^2+||\mathbf{w}_{22}||^2\\
           s.t. \quad\quad  &\log(1+\text{SINR}_{0\rightarrow0})\ge r_0,
  \\& R_j\ge r_j,~j=1,2
  \\
  &                ||\mathbf{w}_{10}||^2+||\mathbf{w}_{11}||^2\le P_{max},
  \\
  &                 ||\mathbf{w}_{20}||^2+||\mathbf{w}_{22}||^2\le P_{max}.
}

Compared with problem (\ref{Ch1_P1}), there are two fewer constraints in problem (\ref{Ch1_DPC1}). Thus, the minimal transmission power of DPC provides a lower bound of the minimal transmission power of N-NOMA. However, due to the high complexity, it is difficult to implement DPC in practical systems. Moreover, superposition coding used in N-NOMA is much easier to implement than DPC. Thus, an interesting question is that whether N-NOMA can achieve the same minimal transmission power as DPC under some channel conditions. The channel makes this true is called quasi-degraded channel of N-NOMA.
\begin{Definition}[Qusai-degraded channel for N-NOMA]
Assume the decoding order of N-NOMA is that CoMP user is prior to NOMA user, and the encoding order of DPC
is that CoMP user is prior to NOMA user, given the target rates $r_j$ ($j=0,1,2$) and the maximum power of each BS $P_{max}$,  the channels $\{\mathbf{h}_{10},\mathbf{h}_{11},\mathbf{h}_{20},\mathbf{h}_{22}\}$ in N-NOMA are quasi-degraded if and only if the minimal transmission powers of N-NOMA and DPC are the same.
\end{Definition}

\section{Quasi-degraded channel condition}
In the following, we would like the derive the expression for the condition that the quasi-degraded channel should meet, which is called QD condition\cite{chen2016application}.
To this end, it is necessary to first obtain the closed-form optimal solution of problem (\ref{Ch1_DPC1}).

Without loss of generality, it is assumed that $||\mathbf{h}_{20}||>||\mathbf{h}_{10}||$. It can be found that the optimal solution of problem (\ref{Ch1_DPC1}) subjects to a specific format as shown in the following lemma.
\begin{Lemma}\label{lemmma_format}
The optimal solution of problem (\ref{Ch1_DPC1}) can be expressed as follows:
\Equ{
&\mathbf{w}_{10}^{D*}=\sqrt{P_{10}}\mathbf{h}_{10}/||\mathbf{h}_{10}||
\\
&\mathbf{w}_{20}^{D*}=\sqrt{P_{20}}\mathbf{h}_{20}/||\mathbf{h}_{20}||
\\
&\mathbf{w}_{11}^{D*}(x)=\sqrt{P_{11}(x)}\frac{\left( \mathbf{I}-x\mathbf{h}_{10}\mathbf{h}_{10}^{H}/||\mathbf{h}_{10}||^2 \right) \mathbf{h}_{11}}{||\left( \mathbf{I}-x\mathbf{h}_{10}\mathbf{h}_{10}^{H}/||\mathbf{h}_{10}||^2 \right) \mathbf{h}_{11}||},\\
& P_{11}(x)=\frac{\sigma ^2\epsilon _1\left( ||\mathbf{h}_{11}||^2-\left( 2x-x^2 \right) ||\mathbf{h}_{10}\mathbf{h}_{11}^{H}||^2/||\mathbf{h}_{10}||^2 \right)}{\left( ||\mathbf{h}_{11}||^2-x||\mathbf{h}_{10}\mathbf{h}_{11}^{H}||^2/||\mathbf{h}_{10}||^2 \right) ^2},
\\
&\mathbf{w}_{22}^{D*}(y)=\sqrt{P_{22}(y)}\frac{\left( \mathbf{I}-y\mathbf{h}_{20}\mathbf{h}_{20}^{H}/||\mathbf{h}_{20}||^2 \right) \mathbf{h}_{22}}{||\left( \mathbf{I}-y\mathbf{h}_{20}\mathbf{h}_{20}^{H}/||\mathbf{h}_{20}||^2 \right) \mathbf{h}_{22}||},\\
&P_{22}(y)=\frac{\sigma^2\epsilon _2\left( ||\mathbf{h}_{22}||^2-\left(2y-y^2 \right) ||\mathbf{h}_{20}\mathbf{h}_{22}^{H}||^2/||\mathbf{h}_{20}||^2 \right)}{\left(||\mathbf{h}_{22}||^2-y||\mathbf{h}_{20}\mathbf{h}_{22}^{H}||^2/||\mathbf{h}_{20}||^2 \right)^2}.
}
where $0<y\leq\epsilon_0/\left( 1+\epsilon_0 \right)$, $0<x\le \epsilon_0/\left( 1+\epsilon_0 \right)$ are  undetermined coefficients which will be given later.
\end{Lemma}
\begin{IEEEproof}
Please refer to Appendix A.
\end{IEEEproof}

In Lemma $1$, there are four undetermined variables, i.e., $P_{10}$, $P_{20}$, $x$, and $y$. Next, we need to find the optimal $P_{10}$, $P_{20}$, $x$, and $y$,  which are denoted by $P_{10,opt}$, $P_{20,opt}$, $x_{opt}$ and $y_{opt}$, respectively. An interesting observation is that the forms of the optimal precoding vectors of N-NOMA is similar to that of single cell NOMA as in \cite{chen2016application}. However, the optimal choice of $x$ ($y$) is different from and more complicated than that in single cell NOMA.

According to the values of $P_{10}$ and $P_{20}$, the solution of (\ref{Ch1_DPC1}) can be classified into three cases:
\begin{itemize}
\item Case I: $P_{20}>0$ and $P_{10}=0$;
\item Case II: $P_{20}>0$ and $P_{10}>0$;
\item Case III: $P_{20}=0$ and $P_{10}>0$.
\end{itemize}

In the following, the optimal solution for each case will be given first, and the optimal solutions of  three cases will then be compared to determine the global optimal solution of problem (\ref{Ch1_DPC1}).

For notational simplicity, define the following functions:
\Equ{
&F_{20}\left( x \right) =\frac{\sigma ^2\epsilon _0+\epsilon _0||\mathbf{h}_{20}^{H}\hat{\mathbf{w}}_{22}||^2}{||\mathbf{h}_{20}||^2}+\frac{\sigma ^2\epsilon _0\epsilon _1||\mathbf{h}_{10}^{H}\mathbf{h}_{11}||^2\left( 1-x \right) ^2}{||\mathbf{h}_{20}||^2\left( A-Bx \right)^2},
\\
&F_{10}\left( y \right) =\frac{\sigma ^2\epsilon _0+\epsilon _0||\mathbf{h}_{10}^{H}\hat{\mathbf{w}}_{11}||^2}{||\mathbf{h}_{10}||^2}+\frac{\sigma ^2\epsilon _0\epsilon _2||\mathbf{h}_{20}^{H}\mathbf{h}_{22}||^2\left( 1-y \right) ^2}{||\mathbf{h}_{10}||^2\left( C-Dy \right)^2},
}
where $A=||\mathbf{h}_{11}||^2$, $B=||\mathbf{h}_{10}^{H}\mathbf{h}_{11}||^2/||\mathbf{h}_{10}||^2$,
$C=||\mathbf{h}_{22}||^2$, $D=||\mathbf{h}_{20}^{H}\mathbf{h}_{22}||^2/||\mathbf{h}_{20}||^2$,
$\hat{\mathbf{w}}_{22}=\mathbf{w}^{D^*}_{22}\left(\epsilon_0/(1+\epsilon_0)\right)$,
and $\hat{\mathbf{w}}_{11}=\mathbf{w}^{D^*}_{11}\left(\epsilon_0/(1+\epsilon_0)\right)$,

In the following, the optimal solution for  $P_{20}>0$ and $P_{10}=0$ is firstly provided.

\begin{Theorem}\label{theorem2_Ch1}
Given $||\mathbf{h}_{20}||>||\mathbf{h}_{10}||$, problem (\ref{Ch1_DPC1}) is feasible under the case where $P_{20}>0$ and $P_{10}=0$ if and only if:
\Equ{\label{Ch1_opt_con1}
  P_{11}\left( 0 \right) <P_{max} \text{ and } F_{20}\left( \tilde{x}_B \right) \le P_{max}-P_{22}(\epsilon_0/(1+\epsilon_0))}
where,
\Equ{
  \tilde{x}_B=\begin{cases}
    \frac{\epsilon_0}{1+\epsilon_0}, \text{ if}\,\,P_{11}\left( \frac{\epsilon_0}{1+\epsilon_0} \right) \le P_{max}\\
    A/2B, \text{ else if}\,\, B=\frac{\sigma ^2\epsilon _1}{P_{max}}\\
    \frac{B-ABP_{max}/\sigma ^2\epsilon _1+\sqrt{B\left( A-B \right) \left( AP_{max}/\sigma ^2\epsilon _1-1 \right)}}{B-B^2P_{max}/\sigma ^2\epsilon _1},\text{ otherwise}.\\
  \end{cases}
}
If the above condition holds, the optimal solution of problem (\ref{Ch1_DPC1}) for the case with $P_{20}>0$ and $P_{10}=0$ can be given by:
\Equ{
&P_{10,opt}=0,\\
&P_{20,opt}=F_{20}\left( x_{opt} \right),\\
&y_{opt}=\frac{\epsilon_0}{1+\epsilon_0},\\
&x_{opt}=\begin{cases}
	\tilde{x}_B, ~\tilde{x}_B<\frac{\epsilon_0}{\epsilon_0+||\mathbf{h}_{20}||^2/||\mathbf{h}_{10}||^2}\\
	\tilde{x}_A, ~\tilde{x}_A\ge \frac{\epsilon_0}{\epsilon_0+||\mathbf{h}_{20}||^2/||\mathbf{h}_{10}||^2}\\
	\frac{\epsilon_0}{\epsilon_0+||\mathbf{h}_{20}||^2/||\mathbf{h}_{10}||^2}, ~\text{else}\\
\end{cases}
}
where,
\Equ{
&\tilde{x}_A=\begin{cases}
	\frac{A\sqrt{P_A}-1}{B\sqrt{P_A}-1},\text{if}\,\,A\sqrt{P_A}<1\\
	0, \text{else}\\
\end{cases}
\\
&P_A=\left( P_{max}-P_{22}(\epsilon_0/(1+\epsilon_0))-\frac{\sigma ^2\epsilon _0+\epsilon _0||\mathbf{h}_{20}^{H}\mathbf{w}_{22}||^2}{||\mathbf{h}_{20}||^2} \right) \frac{||\mathbf{h}_{20}||^2}{\sigma ^2\epsilon _0\epsilon _1||\mathbf{h}_{10}^{H}\mathbf{h}_{11}||^2}.
}
\end{Theorem}
\begin{IEEEproof}
Please refer to Appendix B.
\end{IEEEproof}

\begin{Corollary}
Given $||\mathbf{h}_{20}||>||\mathbf{h}_{10}||$, if problem (\ref{Ch1_DPC1}) has no power constraint, i.e., $P_{max}=\infty$, then the optimal solution of problem (\ref{Ch1_DPC1}) can be expressed as:
\Equ{
&P_{10,opt}=0,\\
&P_{20,opt}=F_{20}\left( x_{opt} \right),\\
&y_{opt}=\frac{\epsilon_0}{1+\epsilon_0},\\
&x_{opt}=\frac{\epsilon_0}{\epsilon_0+||\mathbf{h}_{20}||^2/||\mathbf{h}_{10}||^2}.
}
\end{Corollary}

The following theorem provides the optimal solution for $P_{10}>0$ and $P_{20}>0$.
\begin{theorem}\label{theorem4_Ch1}
Given $||\mathbf{h}_{20}||>||\mathbf{h}_{10}||$, problem is feasible under the case where $P_{10}>0$ and $P_{20}>0$ if and only if:
\Equ{\label{CaseII_condition}
\begin{cases}
P_{max}\!-\!P_{11}\left( \frac{\epsilon _0}{1+\epsilon _0} \right) >0,
P_{max}\!-\!P_{22}(\epsilon_0/(1+\epsilon_0))>0,\\
||\mathbf{h}_{10}||^2\left(\!P_{max}\!-\!P_{11}\left(\!\frac{\epsilon _0}{1+\epsilon _0}\right)\right) \!+\!||\mathbf{h}_{20}||^2\left(\!P_{max}\!-\!P_{22}(\epsilon_0/(1\!+\!\epsilon_0))\!\right) \ge F_{20}\left( \frac{\epsilon _0}{1+\epsilon _0} \right) ||\mathbf{h}_{20}||^2.
\end{cases}
}
If the above conditions holds, then the optimal solution of problem (\ref{Ch1_DPC1}) can be expressed as:
\Equ{
&P_{10,opt}=F_{20}\left( \frac{\epsilon _0}{1+\epsilon _0} \right) ||\mathbf{h}_{20}||^2/||\mathbf{h}_{10}||^2-||\mathbf{h}_{20}||^2/||\mathbf{h}_{10}||^2\left( P_{max}-P_{22}(\epsilon_0/(1+\epsilon_0)) \right),
\\
&P_{20,opt}=P_{max}-P_{22}(\epsilon_0/(1+\epsilon_0)),
\\
&x_{opt}=\frac{\epsilon _0}{1+\epsilon _0},\\
&y_{opt}=\frac{\epsilon _0}{1+\epsilon _0}.
}
\end{theorem}
\begin{IEEEproof}
Please refer to Appendix C.
\end{IEEEproof}

From Theorem $1$ and $2$, a necessary condition for the feasibility under the first two cases can be easily obtained, as highlighted in the following.
\begin{Corollary}
When $P_{20}>0$, problem (\ref{Ch1_DPC1}) is feasible only if:
\begin{align}
 P_{22}(\epsilon_0/(1+\epsilon_0))<P_{max}.
\end{align}
\end{Corollary}

The optimal solution for $P_{10}>0$ and $P_{20}=0$ is highlighted in the following theorem.
\begin{Theorem}
$||\mathbf{h}_{20}||>||\mathbf{h}_{10}||$, problem is feasible under the case where $P_{10}>0$ and $P_{20}=0$ if and only if:
\begin{align}\label{Fea_CaseIII}
 P_{22}(0)<P_{max} \text{ and } F_{10}(\tilde{y}_B)\leq P_{max}-P_{11}(\epsilon_0/(1+\epsilon_0)),
\end{align}
where
\Equ{
  \tilde{y}_B=\begin{cases}
    \frac{\epsilon_0}{1+\epsilon_0}, \text{ if}\,\,P_{22}\left( \frac{\epsilon_0}{1+\epsilon_0} \right) \le P_{max}\\
    C/2D, \text{ else if}\,\, D=\frac{\sigma ^2\epsilon _2}{P_{max}}\\
    \frac{D-CDP_{max}/\sigma ^2\epsilon _2+\sqrt{D\left( C-D \right) \left( CP_{max}/\sigma ^2\epsilon _2-1 \right)}}{D-D^2P_{max}/\sigma ^2\epsilon _2},\text{ otherwise}.\\
  \end{cases}
}
If (\ref{Fea_CaseIII}) is satisfied, the optimal solution of problem (\ref{Ch1_DPC1}) for the case with $P_{10}>0$ and $P_{20}=0$ can be given by:
\begin{align}
 &P_{10,opt}=F_{10}(y_{opt}),\\
&P_{20,opt}=0\\
&x_{opt}=\frac{\epsilon_0}{1+\epsilon_0},\\
&y_{opt}=\tilde{y}_B.
\end{align}
\end{Theorem}
\begin{IEEEproof}
Please refer to Appendix D.
\end{IEEEproof}

Until now, the optimal solution for each case is obtained. Obviously, directly comparing the optimal values of the three cases, the optimal solution for problem (\ref{Ch1_DPC1}) can be obtained. However, this method provides little insight of the problem. Indeed, there is a more elegant way to determine the optimal solution of
problem (\ref{Ch1_DPC1}), which can provides more insight. Interestingly, the optimal solutions of the three cases have priorities as highlighted in the following.
\begin{Proposition}
If problem (\ref{Ch1_DPC1}) is feasible when $P_{20}>0$, then the optimal solution of problem (\ref{Ch1_DPC1}) should satisfy $P_{20}>0$.
\end{Proposition}
\begin{IEEEproof}
Please refer to Appendix E.
\end{IEEEproof}

The above proposition means that if one of the first two cases is feasible, then the optimal solution must be one of the first two cases. In other words, the optimal solutions of the first two cases must be better than that of the third case.

The following corollary, which can be easily obtained from the proof for Proposition $1$, is useful for deciding the optimal solution belongs to which case.
\begin{Corollary}
When $P_{22}(\epsilon_0/(1+\epsilon_0))<P_{max}$, for each feasible solution of case III, there must exist a better solution which belongs to Case I or II.
\end{Corollary}

One important application of the above corollary is that: when $P_{22}(\epsilon_0/(1+\epsilon_0))<P_{max}$ and there is no feasible solution of Case I nor II, it can be concluded that problem (\ref{Ch1_DPC1}) is infeasible.

Further, the priority between the optimal solutions of the first two cases can also be determined, as described in the following proposition.
\begin{Proposition}\label{Pro2}
If condition (\ref{Ch1_opt_con1}) holds, then there is no feasible solution of Case II which is better than the optimal solution of Case I.
\end{Proposition}
\begin{IEEEproof}
Please refer to Appendix F.
\end{IEEEproof}

 \begin{algorithm}[!t]
  \caption{Optimal solution of problem (\ref{Ch1_DPC1})}
  \label{algorithm_DPC}
   \KwIn{$\mathbf{h}_{10}$, $\mathbf{h}_{11}$, $\mathbf{h}_{20}$, $\mathbf{h}_{22}$, $r_j$ ($j=1,2,3$), $P_{max}$ }
   \KwOut{$\mathbf{w}_{10}^{D*}$, $\mathbf{w}_{11}^{D*}$, $\mathbf{w}_{20}^{D*}$, $\mathbf{w}_{22}^{D*}$}
   \eIf{$P_{22}(\epsilon_0/(1+\epsilon_0))<P_{max}$}{
     \lIf{(\ref{Ch1_opt_con1}) holds}{
     Calculate the optimal solution according to Theorem $1$\;
     }
     \lElseIf{(\ref{CaseII_condition}) holds}
     {Calculate the optimal solution according to Theorem $2$\;}
     \lElse{
     output Infeasible;
     }}
   {
   \lIf{(\ref{Fea_CaseIII}) holds}
     {Calculate the optimal solution according to Theorem $3$\;}
     \lElse{
     output Infeasible;
     }
   }
\end{algorithm}
By concluding the above results, the optimal solution of Problem (\ref{Ch1_DPC1}) can be efficiently obtained by Algorithm \ref{algorithm_DPC}.
And then, according to the relationship between problem  (\ref{Ch1_P1}) and problem  (\ref{Ch1_DPC1}), the quasi-degraded channel condition can be obtained, as shown in the following theorem.
\begin{theorem}
If problem (\ref{Ch1_DPC1}) is feasible, the channels are quasi-degraded if and only if:  there exists one pair of  optimal solution of problem (\ref{Ch1_DPC1}), denoted by $\left\{\mathbf{w}_{10}^{D*},
\mathbf{w}_{11}^{D*},\mathbf{w}_{20}^{D*},\mathbf{w}_{22}^{D*}\right\}$, satisfying the following conditions:
\Equ{
  \,\,\,   \text{if }\mathbf{w}_{10}^{D*}\ne 0, ~ -||\mathbf{h}_{11}^{H}\mathbf{w}_{10}^{D*}||^2+\epsilon _0||\mathbf{h}_{11}^{H}\mathbf{w}_{11}^{D*}||^2+\epsilon _0\sigma ^2\le 0,
  \\
  \,\,\,   \text{if }\mathbf{w}_{20}^{D*}\ne 0, ~-||\mathbf{h}_{22}^{H}\mathbf{w}_{20}^{D*}||^2+\epsilon _0||\mathbf{h}_{22}^{H}\mathbf{w}_{22}^{D*}||^2+\epsilon _0\sigma ^2\le 0.
}
\end{theorem}

\section{Application of quasi-degraded channels}
\subsection{H-N-NOMA}
Note that, according to the previous discussion, N-NOMA can achieve the same performance as the DPC scheme only when the channels are degraded. Thus, when the channels are not degraded, there is performance loss by applying N-NOMA compared to DPC. Besides, when the channels are degraded, closed-form expressions for the optimal precoding vectors of N-NOMA can be obtained by using algorithm \ref{algorithm_DPC}, which are the same as those of DPC. However,  closed-form expressions are not available when the channels are not degraded and the acquisition of precoding vectors relies on iterative algorithms, which is not efficient. Thus, in this paper, to reduce the precoding complexity, a novel H-N-NOMA scheme as shown in algorithm \ref{algorithm_H_N_NOMA} is proposed. In the proposed H-N-NOMA scheme, N-NOMA is adopted if the channels are degraded, otherwise, ZFBF is adopted. The superiority of H-N-NOMA will be shown in Section V. The advantage of H-N-NOMA can be further exploited by combining with QDUP, as described in the next subsection.
 \begin{algorithm}[!t]
  \caption{H-N-NOMA scheme}
  \label{algorithm_H_N_NOMA}
      \KwIn{$\mathbf{h}_{10}$, $\mathbf{h}_{11}$, $\mathbf{h}_{20}$, $\mathbf{h}_{22}$, $r_j$ ($j=1,2,3$), $P_{max}$ }
   \KwOut{transmisson strategy}
 \eIf{quasi degraded channel}{
  choose N-NOMA transmisson\;
  }{
  choose ZFBF transmission\;
  }
\end{algorithm}
\begin{algorithm}[!t]
   \vspace{0em}
\setlength{\abovecaptionskip}{0em}   
\setlength{\belowcaptionskip}{-2em}   
  \caption{Qusai-degradation based user pairing (QDUP)}
  \label{algorithm_QDUP}
   \KwIn{channel informmation, $K$, $P_{max}$}
   \KwOut{$\pi_1(k)$, $\pi_2(k)$, $S(k)$}\tcp{$S(k)=1$, choose N-NOMA; $S(k)=0$, choose ZFBF}
 \For{k=1:K}{
   $\text{Flag}=0$\;
   \For{i=1:K}{
     \For{j=1:K}{
       \If{$U_{1,i}$ and $U_{2,j}$ haven't been paired}{
         \If{$U_{1,i}$, $U_{2,j}$ and $U_{0,k}$ have quasi-degraded channels}{
               $\pi_1(k)=i$; $\pi_2(k)=j$; $S(k)=1$;
               $\text{Flag}=1$; break\;
         }
         \If{$U_{1,i}$, $U_{2,j}$ and $U_{0,k}$ have orthogonal channels}{
               $\pi_1(k)=i$; $\pi_2(k)=j$; $S(k)=0$;
               $\text{Flag}=1$;
               break;
         }
       }
     }
   }
   \If{$\text{Flag}=0$}{
     find $\pi_1(k)=\text{argmin}\frac{||(\mathbf{h}_{11}^i)^H\mathbf{h}_{10}^k||^2}{||(\mathbf{h}_{11}^i)||^2||\mathbf{h}_{10}^k||^2}$;
     find $\pi_2(k)=\text{argmin}\frac{||(\mathbf{h}_{22}^j)^H\mathbf{h}_{20}^k||^2}{||(\mathbf{h}_{22}^i)||^2||\mathbf{h}_{20}^k||^2}$\;
     $S(k)=0$;
   }
 }
\end{algorithm}

\subsection{Quasi-degradation based user pairing (QDUP) for multiple N-NOMA groups}
\begin{figure}[!t]
\vspace{0em}
\setlength{\abovecaptionskip}{0em}   
\setlength{\belowcaptionskip}{-1em}   
\centering
\includegraphics[width=3.5in]{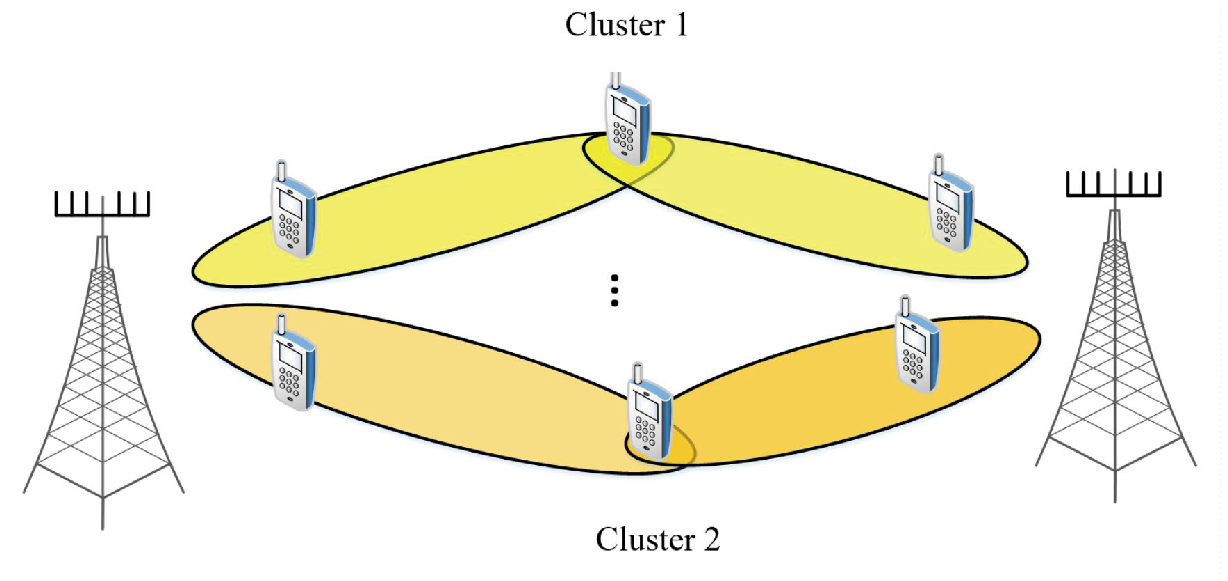}
\caption{Illustration of the system model for for multiple N-NOMA groups.}
\label{system_model2}
\end{figure}
Consider a scenario with multiple CoMP users as shown in Fig. \ref{system_model2}. There are $3K$ users, including: (a) $K$ CoMP users, denoted by $U_{0,k}$, $1\leq k \leq K$; (b) $K$ NOMA
 users which are near to BS $1$, denoted by $U_{1,i}$, $1\leq i \leq K$; (c) $K$ NOMA
 users which are near to BS $2$, denoted by $U_{2,j}$, $1\leq j \leq K$. The channel between $U_{1,i}$ and BS $1$ is denoted by $\mathbf{h}^{i}_{11}$, the channel between $U_{2,j}$ and BS $2$ is denoted by $\mathbf{h}^{j}_{22}$,and the channels between $U_{0,k}$ and BS $1$ and $2$ are denoted by $\mathbf{h}^{k}_{10}$ and $\mathbf{h}^{k}_{20}$, respectively.

The $3K$ users are divided into $K$ groups. Each group consists of a CoMP user, a NOMA user of BS $1$ and a NOMA user of BS $2$. The index of the NOMA user of BS $1$ which is paired with CoMP user $U_{0,k}$ is denoted by $\pi_{1}(k)$, and the index of the NOMA user of BS $2$ which is paired with CoMP user $U_{0,k}$ is denoted by $\pi_{2}(k)$. In this paper, TDMA is applied to serve different groups, i.e., each group is allocated with an independent time slot, yielding no interference between different groups. Besides, equal power allocation is considered for each group, i.e., the largest power allocated to a group by a BS is $P_{max}/K$. In this scenario, the power minimization problem can be formulated as follows:
 \SubEquL{Ch1_P_multi}{
  \underset{\pi _1,\pi _2,\mathbf{w}_{10}^{k},\mathbf{w}_{20}^{k},\mathbf{w}_{11}^{\pi _1(k)},\mathbf{w}_{22}^{\pi _2(k)}}{\min} &\sum_{k=1}^K{\,\,||\mathbf{w}_{10}^{k}||^2+||\mathbf{w}_{20}^{k}||^2+||\mathbf{w}_{11}^{\pi _1(k)}||^2+||\mathbf{w}_{22}^{\pi _2(k)}||^2}
  \\
  \,\,        s.t.  ~~   & R_{j}^{k}\ge r_j,0\le j \le2, 1\le k \le K,
  \\
  \,\,       &              ||\mathbf{w}_{10}^{k}||^2+||\mathbf{w}_{11}^{\pi _1(k)}||^2\le P_{max}/K, 1\le k \le K,
  \\
  \,\,       &              ||\mathbf{w}_{20}^{k}||^2+||\mathbf{w}_{22}^{\pi _2(k)}||^2\le P_{max}/K, 1\le k \le K,
 }
where $R_{j}^k$ is the achievable rate of the user in the $k$-th group, which depends on the specific transmission scheme applied for this group. Obviously, the task is to group user and design precoding vectors for each group.

Problem (\ref{Ch1_P_multi}) is a mixed integer programming problem,  finding its optimal solution is very challenging.
In this paper, a novel transmission scheme termed H-N-NOMA/QDUP is proposed. First, based on quasi-degraded channel condition obtained in the previous section, a greedy user pairing algorithm termed QDUP with low complexity is proposed to provide a sub-optimal solution.
The proposed QDUP preferentially pair users whose channels satisfy QD condition or orthogonal ZFBF,
to make the best N-NOMA and ZFBF. Then, in each group,  H-N-NOMA can be applied to serve users.

\section{Numerical results}
In this section, numerical results are presented to demonstrate the channels' quasi-degradation characteristic of the considered N-NOMA system. The performance of the proposed H-N-NOMA scheme and
H-N-NOMA/QDUP scheme are also presented by providing comparisons with traditional transmission schemes in terms of outage probability and required minimal total transmission power. Rayleigh fading is considered for channel modeling, and the distribution of the channels are as set follows:
$\mathbf{h}_{11}, \mathbf{h}_{22} \sim \mathcal{CN}(0,\sigma_0^2\mathbf{I})$, and
$\mathbf{h}_{10}, \mathbf{h}_{20} \sim \mathcal{CN}(0,\sigma_1^2\mathbf{I})$,
where $\mathcal{CN}$ denotes the circularly symmetric complex gaussian (CSCG) distribution.

\begin{figure*}[!t]
\vspace{-2em}
\setlength{\abovecaptionskip}{0em}   
\setlength{\belowcaptionskip}{-2em}   
\centering
\subfloat[Coverage and QD probability]{\includegraphics[width=0.33\linewidth]{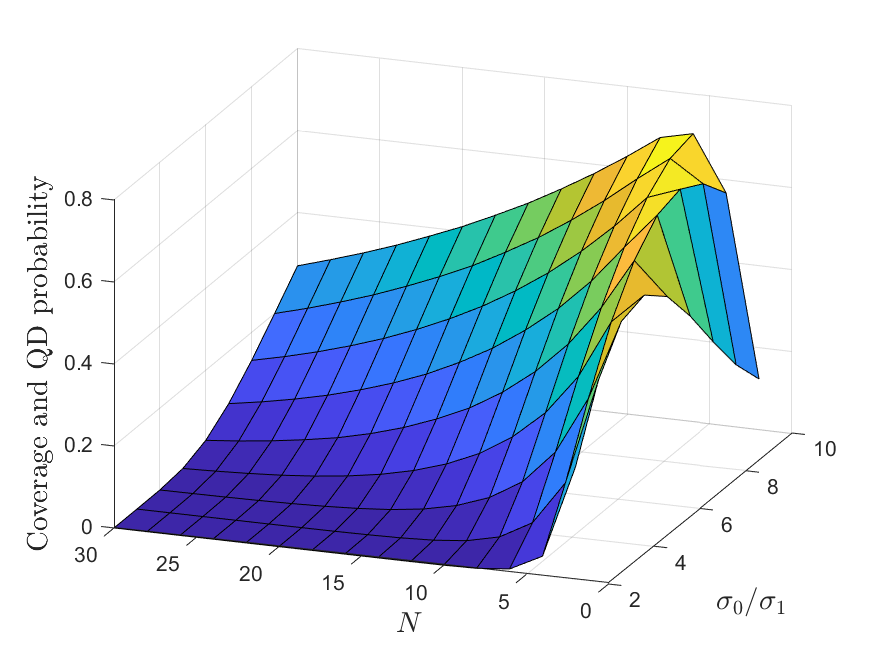}}
\hfil
\subfloat[QD conditioned on coverage]{\includegraphics[width=0.33\linewidth]{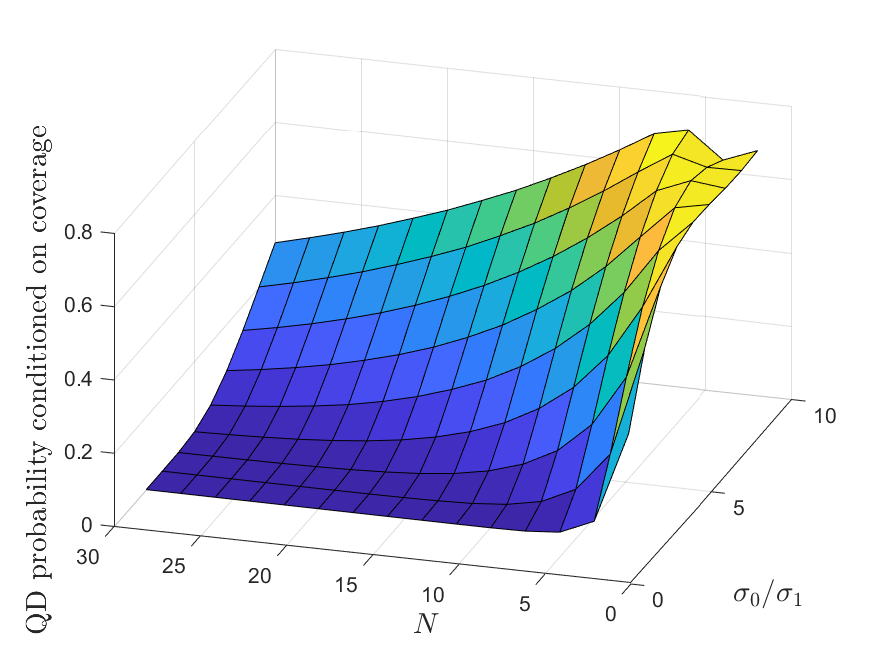}}
\hfil
\subfloat[coverage probability]{\includegraphics[width=0.33\linewidth]{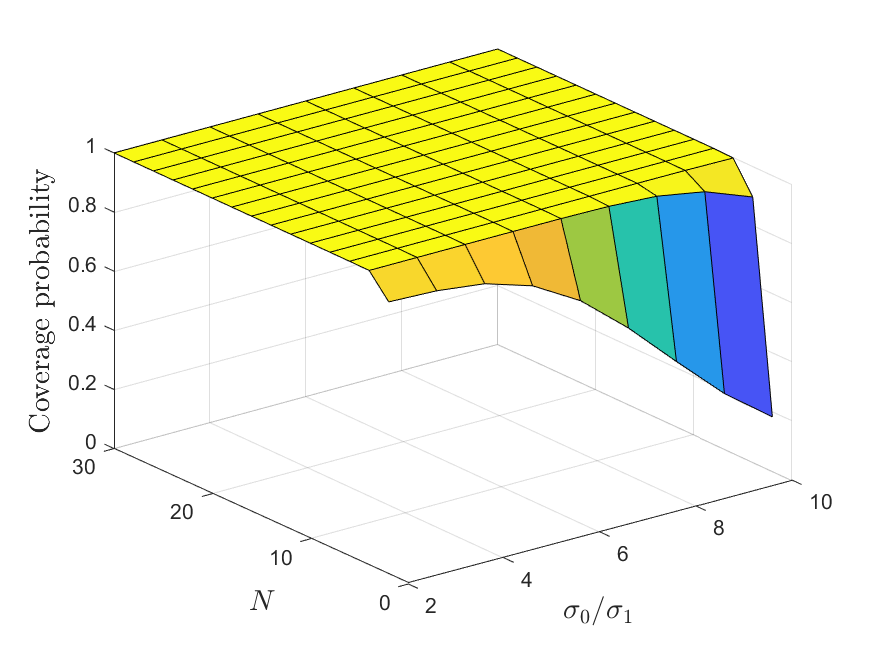}}
\caption{The probabilities relevant to QD. $r_0=0.5$ BPCU,$r_1=2$ BPCU, $r_2=2$ BPCU, $\sigma^2=0.1$, $\sigma_0^2=1$, $P_{max}=1$.}
\label{Probability_QD}
\end{figure*}

Fig. \ref{Probability_QD} shows the probabilities which are relevant to QD versus $N$ and $\sigma_0/\sigma_1$. Note that $\sigma_0/\sigma_1$ represents the disparity between the channel conditions of the CoMP and NOMA users. The larger $\sigma_0/\sigma_1$ is, the weaker the channels of the CoMP user are,
and the larger the disparity is.
Specifically, Fig. \ref{Probability_QD}(a) shows the occurence probability of
channels which make problem (\ref{Ch1_DPC1}) feasible and satisfy QD condition\footnote{Note that coverage means that problem (\ref{Ch1_DPC1}) is feasible.},
Fig. \ref{Probability_QD}(b) shows the conditional QD probability given that problem (\ref{Ch1_DPC1}) is feasible, and Fig. \ref{Probability_QD}(c) shows the probability of that problem (\ref{Ch1_DPC1}) is feasible.
As shown in Fig. \ref{Probability_QD}(c), problem (\ref{Ch1_DPC1}) is more likely to be feasible with a larger $N$ and smaller $\sigma_0/\sigma_1$, which is consistent with the intuition.
From Fig. \ref{Probability_QD}(b), it can be observed that, with large $N$ and small $\sigma_0/\sigma_1$,
the conditional QD probability given problem (\ref{Ch1_DPC1}) is feasible decreases with $N$ and increases with $\sigma_0/\sigma_1$, which is the same as single cell NOMA \cite{chen2016application}. However, the variation becomes a bit complicated when $N$ is small and $\sigma_0/\sigma_1$ is large, which behaves quiet different from single cell NOMA \cite{chen2016application}.
From one point of view, when $N$ is small, the conditional QD probability first increases with $\sigma_0/\sigma_1$ and then decreases slightly.
From another point of view, when $\sigma_0/\sigma_1$ is large, the conditional QD probability will first
decreases, then increases, and finally decreases with $N$.
It is noteworthy that these different behaviors between N-NOMA and single cell NOMA is mainly caused by the
maximum power constraint of each BS added in N-NOMA. Due to this constraint, the optimal solutions switches among three cases for different channel realizations.

\begin{figure*}[!t]
\vspace{-2em}
\setlength{\abovecaptionskip}{0em}   
\setlength{\belowcaptionskip}{-2em}   
\centering
\subfloat[Channels resulting in Case I solution of (\ref{Ch1_DPC1})]{\includegraphics[width=\linewidth]{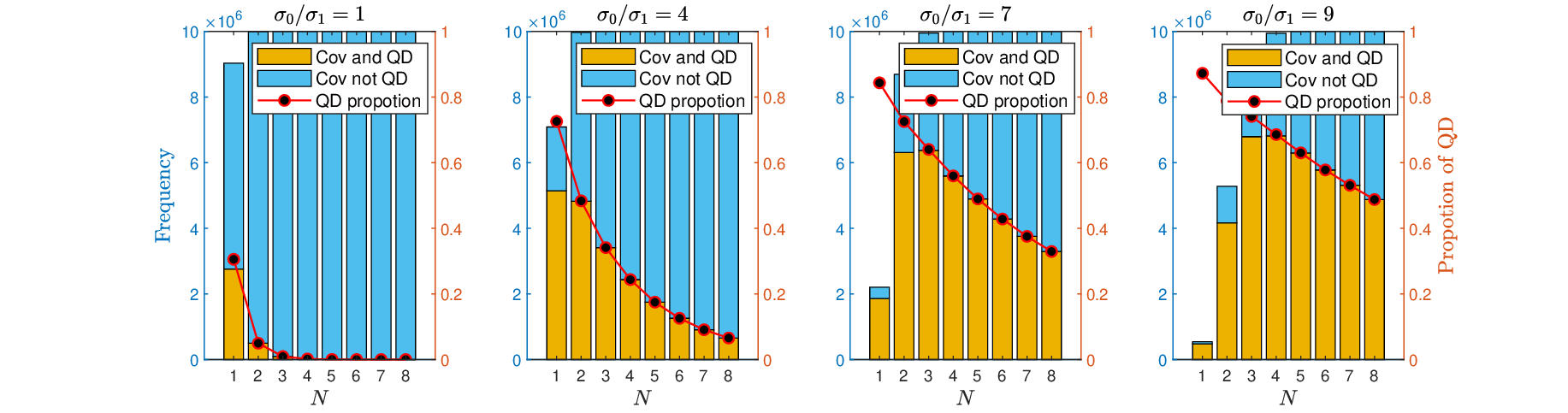}%
\label{bar_DPC_CaseI}}\\
\subfloat[Channels resulting in Case II solution of (\ref{Ch1_DPC1})]{\includegraphics[width=\linewidth]{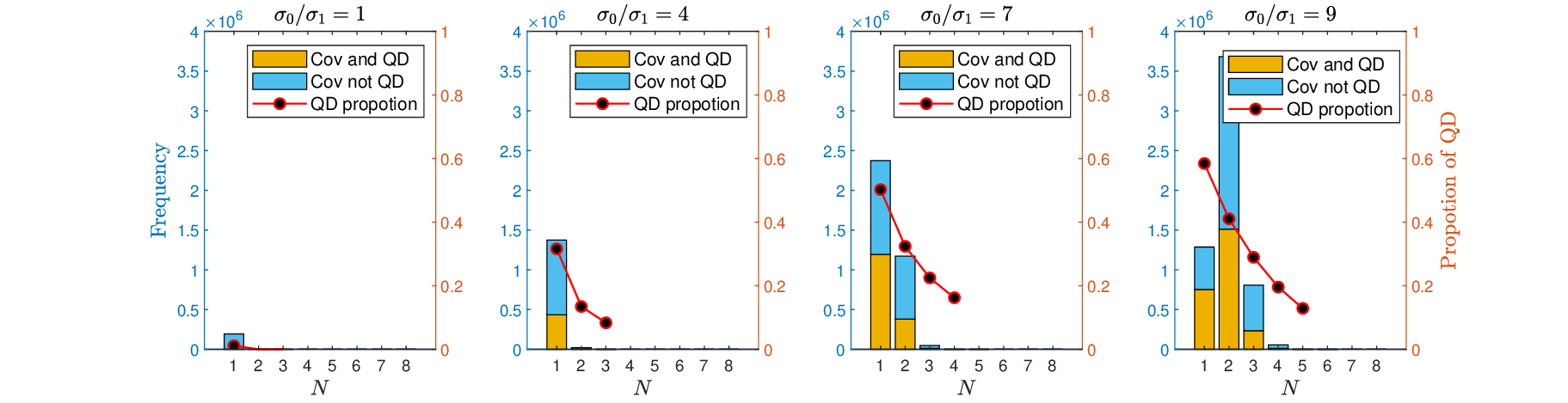}%
\label{bar_DPC_CaseII}}\\
\subfloat[Channels resulting in Case III solution of (\ref{Ch1_DPC1})]{\includegraphics[width=\linewidth]{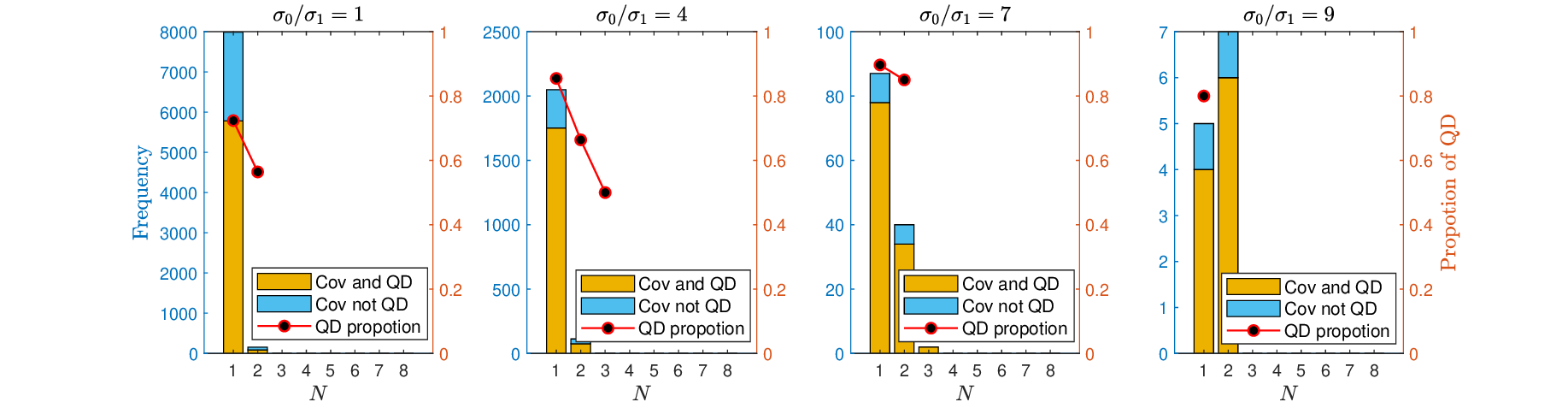}%
\label{bar_DPC_CaseIII}}\\
\caption{A more refined distribution of the channels in $10^7$ realizations. $r_0=0.5$ BPCU, $r_1=2$ BPCU,  $r_2=2$ BPCU, $\sigma^2=0.1$, $\sigma_0^2=1$, $P_{max}=1$.}
\label{dis channels}
\end{figure*}

Fig. \ref{dis channels} shows more detailed statistics obtained from $10^7$ independent channel realizations, which is helpful to understanding the results shown in Fig. \ref{Probability_QD}. Fig. \ref{dis channels}(a), (b) and (c) show the frequency of occurence of the channels that make the optimal solutions of problem (\ref{Ch1_DPC1}) belong to case I, case II and case III, respectively. Channels that are quasi-degraded and not quasi-degraded are also separated in the three sub-figures.
A direct observation from the figure is that the proportion of Case I solutions are dominant compared to Case II and III in most of the situations. And the proportion of Case II solutions is larger than Case III.
Moreover, from Fig. \ref{dis channels}(a), the proportion of channels that result in Case I solution increases with $N$ for fixed $\sigma_0/\sigma_1$, and decreases with $\sigma_0/\sigma_1$ for fixed $N$, when $N$ is not very large. Besides, from Fig. \ref{dis channels}(b) (Fig. \ref{dis channels}(c)), the proportion of the channels that result in Case II (Case III) solution decreases with $N$ when $\sigma_0/\sigma_1=1,4,7$, while that first increases and then increases with $N$ when $\sigma_0\sigma_1=9$.
Another important observations from Fig. \ref{dis channels} are as follows:
\begin{itemize}
 \item when $\sigma_0/\sigma_1$ is fixed, the QD proportion (or conditional QD probability) decreases with $N$.
 \item when $N$ is fixed, the QD proportion (or conditional QD probability) increases with $\sigma_0/\sigma_1$.
\end{itemize}

\begin{figure*}[!t]
\vspace{-2em}
\setlength{\abovecaptionskip}{0em}   
\setlength{\belowcaptionskip}{-1em}   
\centering
\subfloat[$\sigma_1/\sigma_0=\sqrt{20}$,$P_{max}=0.5$]{\includegraphics[width=0.5\linewidth]{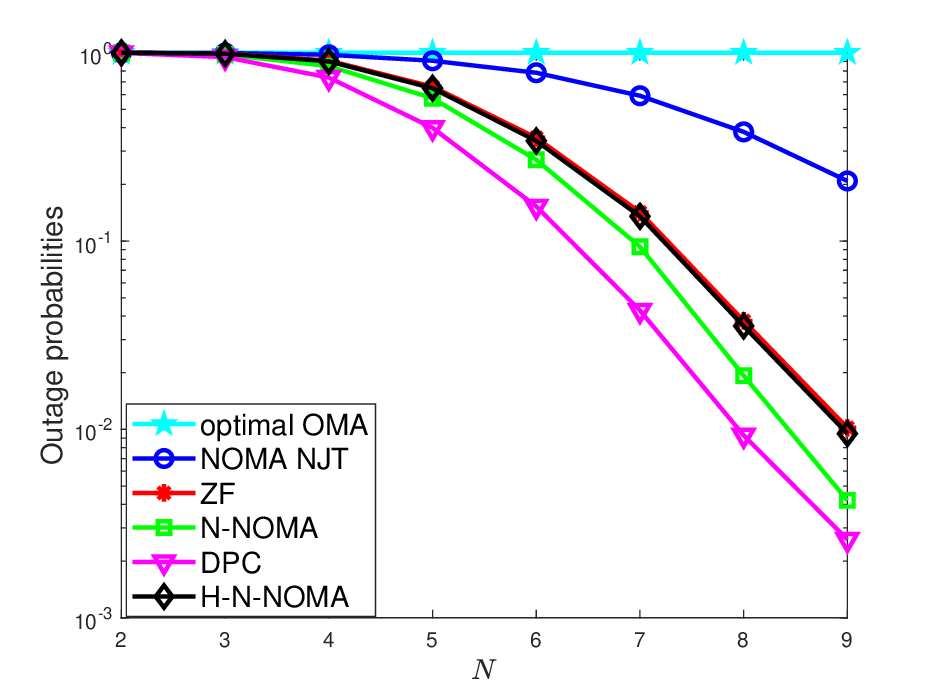}%
\label{single_group_outage_d20P05}}
\hfil
\subfloat[$\sigma_1/\sigma_0=\sqrt{50}$,$P_{max}=1.5$]{\includegraphics[width=0.5\linewidth]{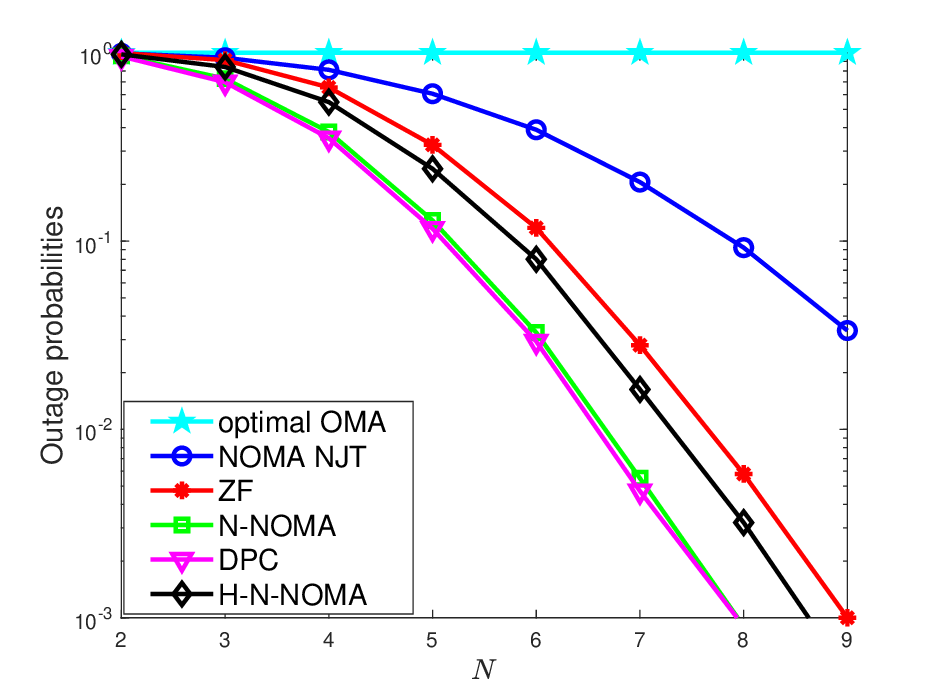}%
\label{single_group_outage_d50P15}}
\caption{Outage probabilities achieved by different schemes for a single group. $r_0=1.5$ BPCU, $r_1=2$ BPCU, $r_2=2$ BPCU, $\sigma^2=0.1$, $\sigma_0^2=1$.}
\label{single_group_outage}
\end{figure*}

\begin{figure*}[!t]
\vspace{-1.5em}
\setlength{\abovecaptionskip}{0em}   
\setlength{\belowcaptionskip}{-2em}   
\centering
\subfloat[$\sigma_1/\sigma_0=\sqrt{20}$]{\includegraphics[width=0.5\linewidth]{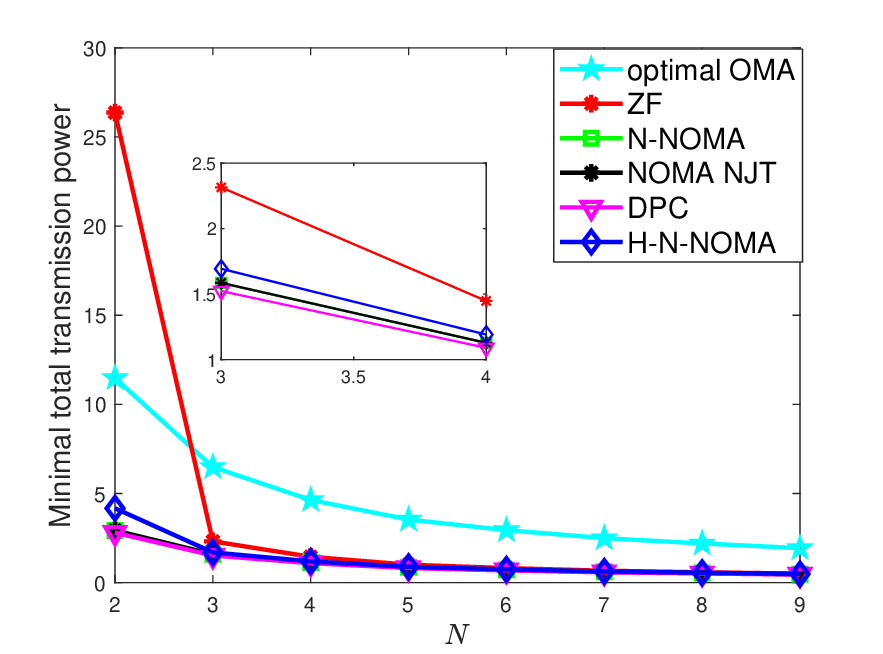}%
\label{single_group_power_d20}}
\hfil
\subfloat[$\sigma_1/\sigma_0=\sqrt{50}$]{\includegraphics[width=0.5\linewidth]{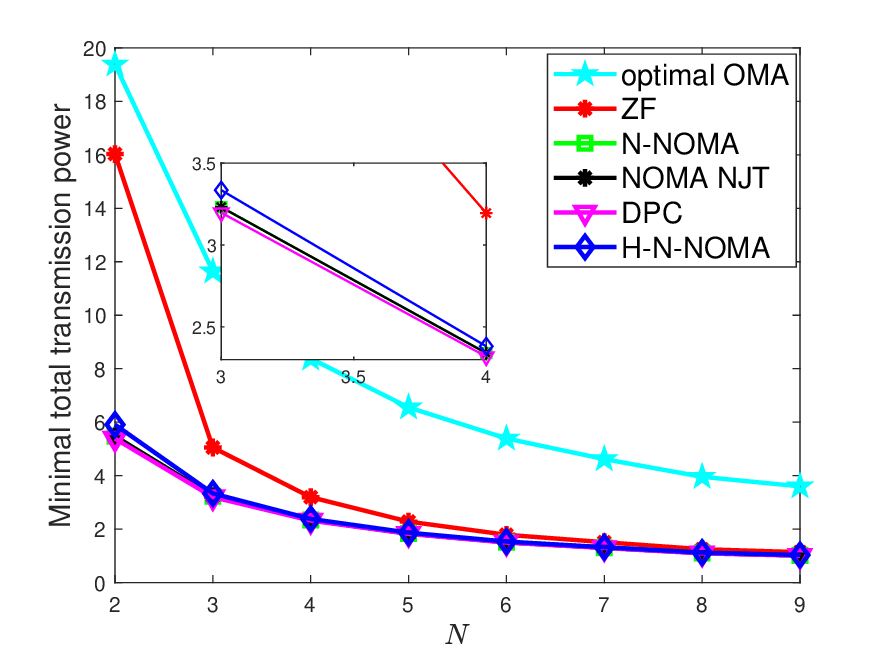}%
\label{single_group_power_d50}}
\caption{Minimal total transmission power required by different schemes for a single group. $P_{max}=\infty$, $r_0=1.5$ BPCU, $r_1=2$ BPCU, $r_2=2$ BPCU, $\sigma^2=0.1$, $\sigma_0^2=1$.}
\label{single_group_power}
\end{figure*}

Figs. \ref{single_group_outage} and \ref{single_group_power} shows the comparison between the proposed H-N-NOMA and traditional schemes, where one CoMP user and two NOMA users are served. Fig.  \ref{single_group_outage} shows the outage probabilities achieved by different schemes versus $N$.
Note that the outage event is defined as that the corresponding total power minimization problem is infeasible.  Fig. \ref{single_group_outage} shows that DPC scheme achieves the best outage performance while optimal TDMA scheme performs worst. It can be shown that H-N-NOMA outperforms optimal OMA, ZFBF and NOMA with NJT schemes. And the gap between H-N-NOMA and ZFBF becomes larger with a larger $\sigma_0/\sigma_1$.
NOMA NJT only performs better than TDMA, which indicates the importance for taking JT into consideration.
Another observation from the figure is that H-N-NOMA achieves higher outage probability compared pure N-NOMA scheme. However, H-N-NOMA has two advantages compared to pure N-NOMA as follows:
\begin{itemize}
 \item Closed-form expressions for the optimal precoding of H-N-NOMA are available. However, it is difficult to obtain the closed-form expressions for N-NOMA, due to the non-convexity of problem (\ref{Ch1_P1}). Thus, the precoder for N-NOMA has to rely on iterative algorithms, which is much more complicated and inefficient for practical implementation than closed-form results driven H-N-NOMA.
 \item As shown later, H-N-NOMA combined with QDUP can outperform N-NOMA with random user pairing. Because efficient user pairing algorithm for N-NOMA is difficult to be obtained, while that for H-N-NOMA can be concisely given through the obtained QD condition.
\end{itemize}

Fig. \ref{single_group_power} shows the minimal total transmission power required by different schemes, when $P_{max}=\infty$. In this figure, an important observation needs to be highlighted that the minimal total transmission power required by NOMA NJT is same as that required by N-NOMA. Which means that N-NOMA degrades to NOMA NJT when the power of one BS is sufficiently large.

\begin{figure*}[!t]
\vspace{-2em}
\setlength{\abovecaptionskip}{0em}   
\setlength{\belowcaptionskip}{-1em}   
\centering
\subfloat[$\sigma_1/\sigma_0=\sqrt{20}$]{\includegraphics[width=0.5\linewidth]{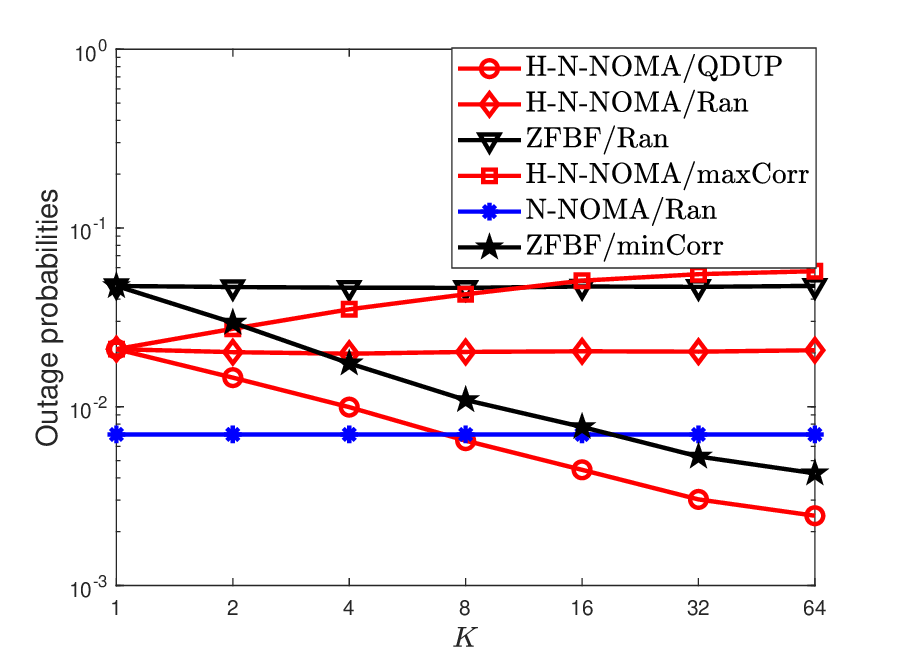}%
\label{mulgroup_N4d20P1_varK}}
\hfil
\subfloat[$\sigma_1/\sigma_0=\sqrt{50}$]{\includegraphics[width=0.5\linewidth]{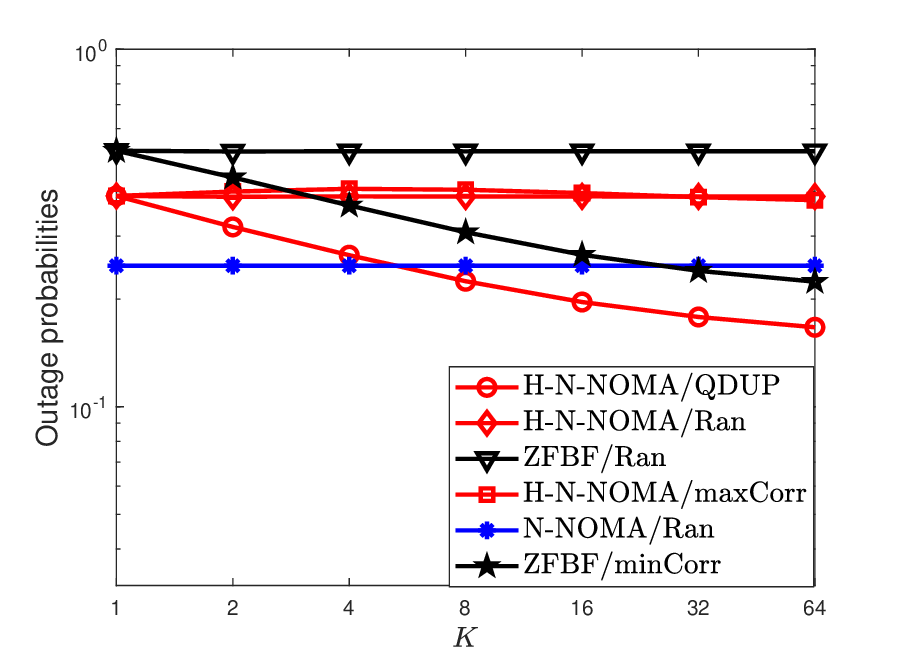}%
\label{mulgroup_N4d50P1_varK}}

\caption{Minimal total transmission power required by different schemes for $K$ groups of users. $P_{max}=1$, $r_0=1$ BPCU,$r_1=2$ BPCU, $r_2=2$ BPCU, $\sigma^2=0.1$, $\sigma_0^2=1$, $N=4$.}
\label{mulgroup_varK}
\end{figure*}

\begin{figure*}[!t]
\vspace{-1.5em}
\setlength{\abovecaptionskip}{0em}   
\setlength{\belowcaptionskip}{-2em}   
\centering
\subfloat[$\sigma_1/\sigma_0=\sqrt{20}$, $P_{max}=0.5$]{\includegraphics[width=0.5\linewidth]{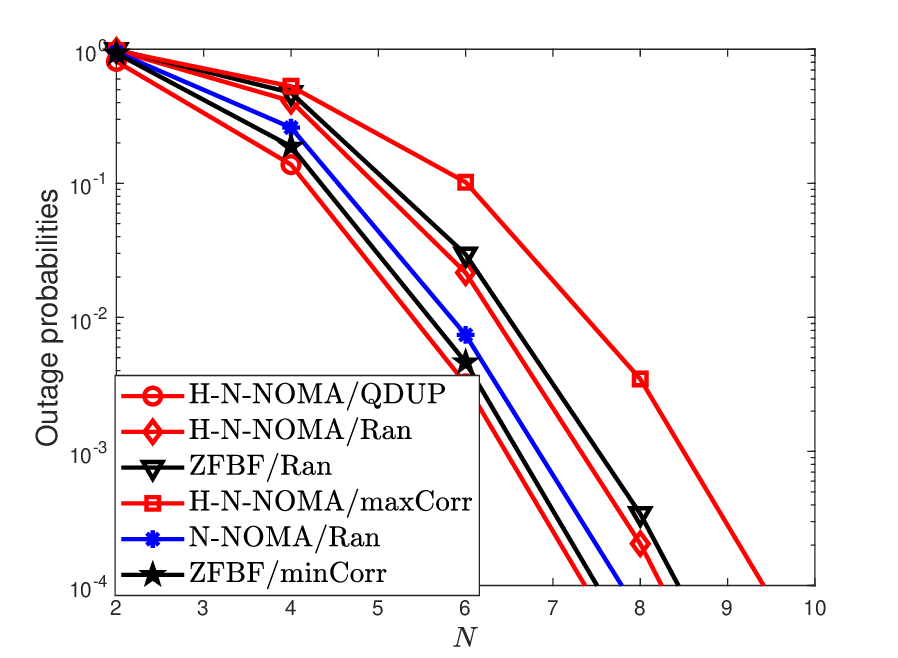}%
\label{mulgro_K32d20P05_varN}}
\hfil
\subfloat[$\sigma_1/\sigma_0=\sqrt{50}$, $P_{max}=1$]{\includegraphics[width=0.5\linewidth]{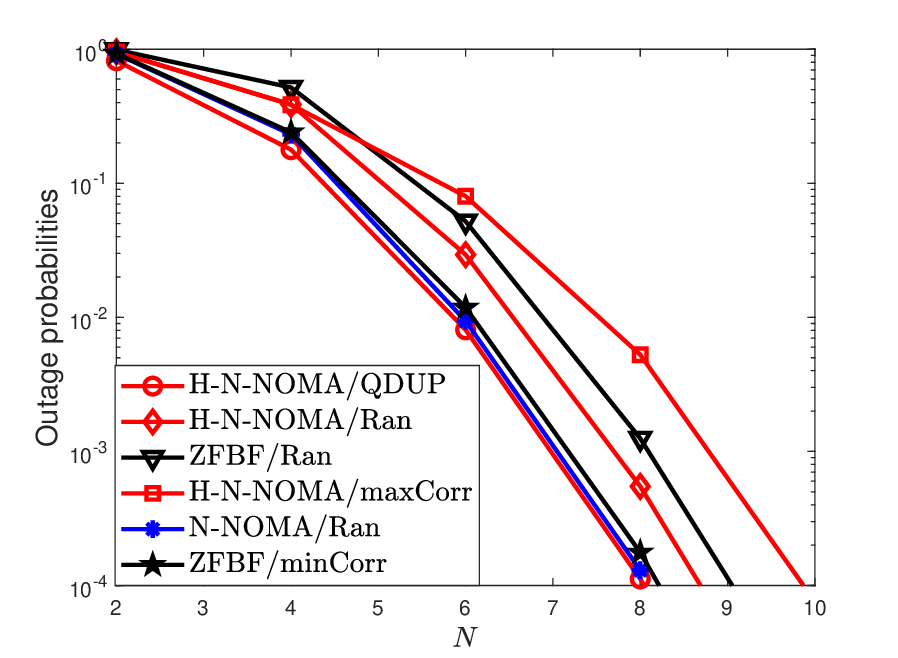}%
\label{mulgro_K32d50P1_varN}}
\caption{Minimal total transmission power required by different schemes for $K$ groups of users. $r_0=1$ BPCU,$r_1=2$ BPCU, $r_2=2$ BPCU, $\sigma^2=0.1$, $\sigma_0^2=1$,$K=32$.}
\label{mulgro_varN}
\end{figure*}

Figs. \ref{mulgroup_varK} and \ref{mulgro_varN} show the comparison among the proposed H-N-NOMA/QDUP algorithm with other benchmark schemes in terms of outage performance, in the scenario with multiple CoMP and NOMA users. These benchmark schemes are:
\begin{itemize}
 \item ZFBF/Ran: users are randomly paired and ZFBF is applied in each group;
  \item ZFBF/minCorr: CoMP users are sequentially paired with the NOMA users with largest channel angels, and ZFBF is applied in each group;
 \item H-N-NOMA/Ran: users are randomly grouped, and H-N-NOMA is applied in each group;
 \item H-N-NOMA/maxCorr: CoMP users are sequentially paired with the NOMA users with smallest channel angels, and H-N-NOMA is applied in each group.
 \item N-NOMA/Ran: users are randomly grouped, and N-NOMA is applied in each group;
\end{itemize}
The outage probabilities in Figs. \ref{mulgroup_varK} and \ref{mulgro_varN} are obtained as follows.
In each trial, users are divided into $K$ groups and for each group, if the target rates of the users cannot be supported by the using the allocated power, then it is counted as one time of outage event. Finally, the outage probability can be calculated as:
\begin{align}
 \frac{\text{number of outage groups}}{K\times \text{number of trials}}.
\end{align}
Fig. \ref{mulgroup_varK} shows that, when $K\geq2$, H-N-NOMA/QDUP outperforms H-N-NOMA/Ran, H-N-NOMA/maxCorr,
ZFBF/Ran and ZFBF/minCorr schemes. When $K$ is small, H-N-NOMA/QDUP achieves higher outage probability than
N-NOMA/Ran. However, the proposed H-N-NOMA/QDUP can effectively exploit the benefit of multi user diversity, and the outage probability decreases with $K$. While the outage probability achieved by N-NOMA/Ran don't vary with $K$. And hence, H-N-NOMA/QDUP gradually outperforms N-NOMA/Ran as $K$ increases. In Fig. \ref{mulgro_varN}, $K$ is set to be $32$, it can be observed that H-N-NOMA/QDUP outperforms all other schemes.
\section{Conclusions}
In this paper, quasi-degraded channel has been studied for downlink N-NOMA system with two BSs.
In quasi-degraded channel, given the users' target rates and the maximum transmit power of each BS,
N-NOMA can achieve the same minimal total transmission power as DPC.
Mathematical condition for channels to satisfy QD has been rigorously derived.
And closed-form expression for the optimal precoding vector of N-NOMA under quasi-degraded channel has also been provided. Based on the obtained closed-form results, a novel H-N-NOMA scheme has been proposed.
Further, for the scenarios with more users, a low-complexity QD based user pairing (QDUP) algorithm has been proposed. Finally,  QDUP and H-N-NOMA have been combined, and it has been shown that the proposed H-N-NOMA/QDUP scheme can effectively exploit the benefit of multi user diversity and outperform existing schemes.
\appendices
\section{Proof for Lemma 1}
Problem (\ref{Ch1_DPC1}) can be rewritten as:
\SubEquL{PCh1_DPC1}{
  \min_{\begin{array}{c}
    \mathbf{w}_{10},\mathbf{w}_{20},
    \mathbf{w}_{11},\mathbf{w}_{22}
  \end{array}} &||\mathbf{w}_{10}||^2+||\mathbf{w}_{20}||^2+||\mathbf{w}_{11}||^2+||\mathbf{w}_{22}||^2\\
    s.t. \quad\quad&-||\mathbf{h}_{10}^{H}\mathbf{w}_{10}||^2-||\mathbf{h}_{20}^{H}\mathbf{w}_{20}||^2\\\notag
            &+\epsilon_0\left(||\mathbf{h}_{10}^{H}\mathbf{w}_{11}||^2+||\mathbf{h}_{20}^{H}\mathbf{w}_{22}||^2+\sigma ^2\right)\le0,
  \\&     -||\mathbf{h}_{11}^{H}\mathbf{w}_{11}||^2+\sigma^2\epsilon_1\le0\\
    &     -||\mathbf{h}_{22}^{H}\mathbf{w}_{22}||^2+\sigma^2\epsilon_2\le0\\
  &                ||\mathbf{w}_{10}||^2+||\mathbf{w}_{11}||^2\le P_{max},
  \\
  &                 ||\mathbf{w}_{20}||^2+||\mathbf{w}_{22}||^2\le P_{max}.
}

This problem is an non-convex problem. However, as shown later, it can be transformed to a convex problem. An interesting observation is that: the directions of the optimal
$\mathbf{w}_{10}$ and $\mathbf{w}_{20}$ should satisfy:
$\tilde{\mathbf{w}}_{10}=\frac{\mathbf{h}_{10}}{||\mathbf{h}_{10}||}$, $\tilde{\mathbf{w}}_{20}=\frac{\mathbf{h}_{20}}{||\mathbf{h}_{20}||}$.
Thus, $\mathbf{w}_{10}$ and $\mathbf{w}_{20}$ can be expressed as:
$\mathbf{w}_{10}=\sqrt{P_{10}}\tilde{\mathbf{w}}_{10}$, $\mathbf{w}_{20}=\sqrt{P_{20}}\tilde{\mathbf{w}}_{20}$.
Then, problem (\ref{PCh1_DPC1}) can be rewritten as:
\SubEquL{PCh1_DPC2}{
  \min_{\begin{array}{c}
    P_{10},P_{20},
    \mathbf{w}_{11},\mathbf{w}_{22}
  \end{array}} &P_{10}+P_{20}+||\mathbf{w}_{11}||^2+||\mathbf{w}_{22}||^2\\
    s.t. \quad\quad&-||\mathbf{h}_{10}||^2P_{10}-||\mathbf{h}_{20}||^2P_{20}\\\notag
            &+\epsilon_0\left(||\mathbf{h}_{10}^{H}\mathbf{w}_{11}||^2+||\mathbf{h}_{20}^{H}\mathbf{w}_{22}||^2+\sigma ^2\right)\le0,
  \\&     -||\mathbf{h}_{11}^{H}\mathbf{w}_{11}||^2+\sigma^2\epsilon_1\le0\\
    &     -||\mathbf{h}_{22}^{H}\mathbf{w}_{22}||^2+\sigma^2\epsilon_2\le0\\
  &                P_{10}+||\mathbf{w}_{11}||^2\le P_{max},
  \\
  &                P_{20}+||\mathbf{w}_{22}||^2\le P_{max},\\
  &                -P_{10}\le0,\\
  &                -P_{20}\le0.
}

It can be found that, the constraints (\ref{PCh1_DPC2}b), (\ref{PCh1_DPC2}e)-(\ref{PCh1_DPC2}h) are convex.
Intuitively, (\ref{PCh1_DPC2}c) is an non-convex constraint, however, in essence,
it is a convex constraint, the reasons are as follows:
first, multiply $\mathbf{w}_{10}$ with a complex number with modulus $1$, denoted by $e^{j\phi}$, such that $\mathbf{h}_{11}^H\mathbf{w}_{10}$ is a positive real number, this operation will not change the optimality of the problem, then   (\ref{PCh1_DPC2}c) can be transformed into a linear constraint, and hence is convex. Similarly, constraint (\ref{PCh1_DPC2}d) is also convex. Thus, problem (\ref{PCh1_DPC2}) is a convex problem.
The Lagrangian of problem (\ref{PCh1_DPC2}) is given by:
\Equ{
   L=&P_{10}+P_{20}+||\mathbf{w}_{11}||^2+||\mathbf{w}_{22}||^2\\\notag
   &+\lambda_1\left(-||\mathbf{h}_{10}||^2P_{10}-||\mathbf{h}_{20}||^2P_{20}
   +\epsilon_0\left(||\mathbf{h}_{10}^{H}\mathbf{w}_{11}||^2+||\mathbf{h}_{20}^{H}\mathbf{w}_{22}||^2+\sigma ^2\right)\right)\\\notag
   &+\lambda_2(-||\mathbf{h}_{11}^{H}\mathbf{w}_{11}||^2
    +\sigma^2\epsilon_1)
    +\lambda_3(-||\mathbf{h}_{22}^{H}\mathbf{w}_{22}||^2+\sigma^2\epsilon_2)\\\notag
 &  +\lambda_4(P_{10}+||\mathbf{w}_{11}||^2-P_{max})
     +   \lambda_5(P_{20}+||\mathbf{w}_{22}||^2-P_{max})
                     -\lambda_6P_{10}
                  -\lambda_7P_{20}.
}
where $\lambda_i\ge0,i=1,2,\cdots,7$ are the Lagrangian multipliers.

It is not difficult to find that the optimal $\lambda_1,\lambda_2$ and $\lambda_3$ must be larger than zero, and equality holds for  the corresponding constraints. Then,
according to the stationarity of the KKT condition, the following relationships can be established:
\begin{align}
    &\frac{\partial{L}}{P_{10}}=1+\lambda_4-\lambda_6-\lambda_1||\mathbf{h}_{10}||^2=0,\\
    &\frac{\partial{L}}{P_{20}}=1+\lambda_5-\lambda_7-\lambda_1||\mathbf{h}_{10}||^2=0,\\
    &\frac{\partial{L}}{\mathbf{w}_{11}}=2(1+\lambda_4)\mathbf{w}_{11}+2\lambda_1\epsilon_0\mathbf{h}_{10}\mathbf{h}_{10}^H\mathbf{w}_{11}
    -2\lambda_2\mathbf{h}_{11}\mathbf{h}_{11}^H\mathbf{w}_{11}=0,\\
    &\frac{\partial{L}}{\mathbf{w}_{22}}=2(1+\lambda_5)\mathbf{w}_{22}+2\lambda_1\epsilon_0\mathbf{h}_{20}\mathbf{h}_{20}^H\mathbf{w}_{22}
    -2\lambda_3\mathbf{h}_{22}\mathbf{h}_{22}^H\mathbf{w}_{22}=0.
\end{align}

From the first two equations above, it is easy to get the following relationships:
\Equ{\label{PCh1_lambda1}
    \frac{\lambda_1}{1+\lambda_4-\lambda_6}=\frac{1}{||\mathbf{h}_{10}||^2},
}
\Equ{\label{PCh1_lambda2}
    \frac{\lambda_1}{1+\lambda_5-\lambda_7}=\frac{1}{||\mathbf{h}_{20}||^2},
}

While from the later two equations, the followings can be obtained:
\Equ{
    \mathbf{w}_{11}=\left(\mathbf{I}+\frac{\lambda_1\epsilon_0}{1+\lambda_4}
    \mathbf{h}_{10}\mathbf{h}_{10}^H
    \right)^{-1}\lambda_2\mathbf{h}_{11}\mathbf{h}_{11}^H\mathbf{w}_{11}
}
\Equ{
    \mathbf{w}_{22}=\left(\mathbf{I}+\frac{\lambda_1\epsilon_0}{1+\lambda_5}
    \mathbf{h}_{20}\mathbf{h}_{20}^H
    \right)^{-1}\lambda_3\mathbf{h}_{22}\mathbf{h}_{22}^H\mathbf{w}_{22}
}

Note that, we can make $\mathbf{h}_{11}^H\mathbf{w}_{11}$ be a positive real number, the direction of $\mathbf{w}_{11}$ can be expressed as:
\Equ{
    \tilde{\mathbf{w}}_{11}=\frac{\left(\mathbf{I}+\frac{\lambda_1\epsilon_0}{1+\lambda_4}
    \mathbf{h}_{10}\mathbf{h}_{10}^H
    \right)^{-1}\mathbf{h}_{11}}{||\left(\mathbf{I}+\frac{\lambda_1\epsilon_0}{1+\lambda_4}
    \mathbf{h}_{10}\mathbf{h}_{10}^H
    \right)^{-1}\mathbf{h}_{11}||}
}

Similarly, the direction of $\mathbf{w}_{22}$ can be expressed as:
\Equ{
    \tilde{\mathbf{w}}_{22}=\frac{\left(\mathbf{I}+\frac{\lambda_1\epsilon_0}{1+\lambda_5}
    \mathbf{h}_{20}\mathbf{h}_{20}^H
    \right)^{-1}\mathbf{h}_{22}}{||\left(\mathbf{I}+\frac{\lambda_1\epsilon_0}{1+\lambda_5}
    \mathbf{h}_{20}\mathbf{h}_{20}^H
    \right)^{-1}\mathbf{h}_{22}||}
}

Further, according to (\ref{PCh1_lambda2}) and Shermon Morrison equation, $\tilde{\mathbf{w}}_{22}$ can be rewritten as:
\Equ{
   \tilde{\mathbf{w}}_{22}=\frac{\left(\mathbf{I}-\frac{\epsilon_0}{1+\epsilon_0}
   \mathbf{h}_{20}\mathbf{h}_{20}^H/||\mathbf{h}_{20}||^2
   \right)\mathbf{h}_{22}}{||\left(\mathbf{I}-\frac{\epsilon_0}{1+\epsilon_0}
   \mathbf{h}_{20}\mathbf{h}_{20}^H/||\mathbf{h}_{20}||^2
   \right)\mathbf{h}_{22}||}
}

let $\mathbf{w}_{22}=\sqrt{P_{22}}\tilde{\mathbf{w}}_{22}$, and let
equality holds in constraint (\ref{PCh1_DPC2}d), the expression of $P_{22}$ can be obtained.
Next, let $t_x=\frac{\lambda_1\epsilon_0}{1+\lambda_4}$ and applying Shermon Morrison equation, it is obtained that:
\begin{align}
    \tilde{\mathbf{w}}_{11}=\frac{\left(\mathbf{I}-\frac{t_x
    \mathbf{h}_{10}\mathbf{h}_{10}^H}{1+t_x||\mathbf{h}_{10}||^2}
    \right)\mathbf{h}_{11}}{||\left(\mathbf{I}-\frac{t_x
    \mathbf{h}_{10}\mathbf{h}_{10}^H}{1+t_x||\mathbf{h}_{10}||^2}
    \right)\mathbf{h}_{11}||}
\end{align}
Further, let
\Equ{\label{Ch1_xt}
x=\frac{t_x||\mathbf{h}_{10}||^2}{1+t_x||\mathbf{h}_{10}||^2},}
and $\mathbf{w}_{11}=\sqrt{P_{11}(x)}\tilde{\mathbf{w}}_{11}$,
and let equality holds in constraint (\ref{PCh1_DPC2}c), the expression of $P_{11}(x)$ can be obtained. Moreover, according to (\ref{PCh1_lambda1}), it is obtained that $0<x\leq\epsilon_0/(1+\epsilon_0)$.

Similarly, $\mathbf{w}_{22}$ can be expressed as:
$\mathbf{w}_{22}=\sqrt{P_{22}(y)}\tilde{\mathbf{w}}_{22}$,
where $y=\frac{t_y||\mathbf{h}_{20}||^2}{1+t_y||\mathbf{h}_{20}||^2}$, $t_y=\frac{\lambda_1\epsilon_0}{1+\lambda_5}$.
And the expression of $P_{22}(y)$ and value range of $y$ can be obtained as same as $P_{11}(y)$ and $x$.
\section{Proof for Theorem 1}
In problem (\ref{PCh1_DPC2}), according to the complementary slackness of the KKT condition, we have:
$\lambda_6P_{10}=0$, $\lambda_7P_{20}=0$.
Since $P_{10}=0$ and $P_{20}>0$ are assumed, it can be concluded that $\lambda_7=0$.
By taking $\lambda_7=0$ into (\ref{PCh1_lambda2}), it is obtained that:
$\frac{\lambda_1}{1+\lambda_5}=\frac{1}{||\mathbf{h}_{20}||^2}$,
Thus $t_y=1/||\mathbf{h}_{20}||^2$ and the value of optimal $y$ denoted by $y_{opt}$ can be obtained as
$y_{opt}=\frac{\epsilon_0}{1+\epsilon_0}$.

Further, by noting that constraint (\ref{PCh1_DPC2}b) should take mark of equality as stated in Appendix B, it is obtained that: $P_{20}=F_{20}(x).$
Based on the above discussions, the primal problem can be simplified to the following optimal problem:
\SubEquL{PCh1_DPCs}{
    \min_{x}~&F_{20}(x)+P_{11}(x)\\
    s.t.~~ &P_{11}(x)\le P_{max},\\
         &F_{20}(x)\le P_{max}-P_{22}(\epsilon_0/(1+\epsilon_0))\\
         &0<x\le\frac{\epsilon_0}{1+\epsilon_0}
}
Thus, the left task is to find the optimal $x$.

Let $G(x)=F_{20}(x)+P_{11}(x)$, and take the derivative of $G(x)$ with respect to $x$, we have
\Equ{
  G'(x)=F'_{20}(x)+F'_{11}(x)
}
where
\Equ{
    F'_{20}(x)=\frac{2\sigma^2\epsilon_0\epsilon_1
        ||\mathbf{h}_{10}^H\mathbf{h}_{11}||^2(A-B)(x-1)}{||\mathbf{h}_{20}||^2(A-Bx)^3}
}
\Equ{
    F'_{11}(x)=\frac{2\sigma^2\epsilon_1(A-B)Bx}{(A-Bx)^3}
}

Thus, in interval $0<x\leq\frac{\epsilon_0}{1+\epsilon_0}$, it is easy to have the following observations:
\begin{itemize}
    \item  $F_{20}$ decreases with $x$;
    \item $P_{11}$ increases with $x$;
    \item $G(x)$ first decreases with $x$ and then increases.
\end{itemize}

Let $G'(x)=0$, the extreme point can be obtained:
\Equ{
    x_{ext}=\frac{\epsilon_0}{\epsilon_0+{||\mathbf{h}_{20}||^2}/{||\mathbf{h}_{10}||^2}}
}

Let $\tilde{x}_B$ be the largest $x$ in interval $(0,{\epsilon_0}/{1+\epsilon_0}]$ which satisfies constraint (\ref{PCh1_DPCs}b), obviously, $\tilde{x}_B$ exists if and only if
$
    P_{11}(0)<P_{max}.
$

If $P_{11}(0)<P_{max}$, then the optimal solution is restricted to be located in $(0,\tilde{x}_{B}]$. Further, take constraint (\ref{PCh1_DPCs}c) into consideration,
let $\tilde{x}_A$ be the minimal $x$ in interval $(0,\tilde{x}_{B}]$ which satisfies constraint (\ref{PCh1_DPCs}c), and $\tilde{x}_A$ exists if and only if:
$
F_{20}(\tilde{x}_B)\le P_{max}-P_{22}(\epsilon_0/(1+\epsilon_0))
$

Under the condition that $\tilde{x}_A$ and $\tilde{x}_B$ exist, it is not to derive their expressions as shown in the theorem.
Hence, the optimal solution of $x$ is restricted to be located in $[\tilde{x}_A,\tilde{x}_B]$. Finally, by using the relationship between $x_{\text{ert}}$ and $\tilde{x}_A$, $\tilde{x}_B$, the expression of the optimal solution can be obtained, and the proof is complete.

\section{Proof for Theorem 2}
Given $P_{10}>0$ and $P_{20}>0$,
according to the complementary slackness of the KKT condition of problem (\ref{PCh1_DPC2}), we have $\lambda_6=0$ and $\lambda_7=0$, and based on (\ref{PCh1_lambda1}) and (\ref{Ch1_xt}), we have
$
 x_{opt}={\epsilon_0}/{1+\epsilon_0}$ and $ y_{opt}={\epsilon_0}/{1+\epsilon_0}.$
Then by applying Lemma $1$, the optimal $P_{11}$, $P_{22}$, $\tilde{\mathbf{w}}_{11}$, and $\tilde{\mathbf{w}}_{22}$ can be determined. Then, the optimization problem (\ref{Ch1_DPC1}) can be simplified to:
\SubEquL{PCh1_DPCs2}{
    \min~ &P_{10}+P_{20}\\
    s.t.~~&||\mathbf{h}_{10}||^2P_{10}+||\mathbf{h}_{20}||^2P_{20}=P_{11}(x_{opt})\\
    & 0\le P_{10}\le P_{max}-P_{11}(x_{opt})\\
        & 0\le P_{20}\le P_{max}-P_{22}(y_{opt}).
}

Problem (\ref{PCh1_DPCs2}) is a linear programming problem with variables $P_{10}$ and $P_{20}$ and it can be easily solved. Constraints (\ref{PCh1_DPCs2}c) and (\ref{PCh1_DPCs2}d) provides the feasibility condition for the problem. Note that $||\mathbf{h}_{20}||$ is larger than $||\mathbf{h}_{10}||$, which means it is better to use as much $P_{20}$ as possible. Based on the observation, the optimal solution can be obtained and the proof is complete.

\section{Proof for Theorem 3}
In problem (\ref{PCh1_DPC2}), according to the complementary slackness of the KKT condition, we have:
$\lambda_6P_{10}=0, \lambda_7P_{20}=0$.
Since $P_{20}=0$ and $P_{10}>0$ are assumed, it can be concluded that $\lambda_6=0$.
By taking $\lambda_6=0$ into (\ref{PCh1_lambda1}), it is obtained that: $
{\lambda_1}/{1+\lambda_4}={1}/{||\mathbf{h}_{10}||^2}$,
Thus $t_x=1/||\mathbf{h}_{10}||^2$ and the value of optimal $x$ denoted by $x_{opt}$ can be obtained:
$x_{opt}=\frac{\epsilon_0}{1+\epsilon_0}$.

Further, by noting that constraint (\ref{PCh1_DPC2}b) should take mark of equality as stated in Appendix B, it is obtained that: $P_{10}=F_{10}(y)$.
Based on the above discussions, the primal problem can be simplified to the following optimal problem:
\SubEquL{PCh1_DPCs3}{
    \min_{x}~&F_{10}(y)+P_{22}(y)\\
    s.t.~~ &P_{22}(y)\le P_{max},\\
         &F_{10}(y)\le P_{max}-P_{11}(\epsilon_0/(1+\epsilon_0))\\
         &0<y\le\frac{\epsilon_0}{1+\epsilon_0}
}

Let $Q(y)=F_{10}(y)+P_{22}(y)$, and take the derivative of $Q(y)$ with respect to $y$, we have
  $Q'(y)=F'_{10}(y)+F'_{22}(y)$,
where
\Equ{
    F'_{10}(y)=\frac{2\sigma^2\epsilon_0\epsilon_2
        ||\mathbf{h}_{20}^H\mathbf{h}_{22}||^2(C-D)(y-1)}{||\mathbf{h}_{10}||^2(C-Dy)^3}
}
\Equ{
    F'_{22}(y)=\frac{2\sigma^2\epsilon_2(C-D)Dy}{(C-Dy)^3}
}

When $0<y<1$, it is not hard to have the following observations:
(a) $F_{10}$ decreases with $y$ and $F'_{10}$ increases with $y$;
(b) $P_{22}$ increases with $y$ and $F'_{22}$ increases with $y$.
Thus $Q'(y)$ increases with $y$ when $0<y<1$.  Let $Q'(y)=0$, the extreme point can be obtained:
\Equ{
    y_{ext}=\frac{\epsilon_0}{\epsilon_0+{||\mathbf{h}_{10}||^2}/{||\mathbf{h}_{20}||^2}},
}
Since $||\mathbf{h}_{10}||<||\mathbf{h}_{20}||$, it can be concluded that $y_{ext}>\epsilon_0/(1+\epsilon_0)$. Thus, in interval  $0<y\leq\frac{\epsilon_0}{1+\epsilon_0}$, $Q(y)$ decreases with $y$.

Let $\tilde{y}_B$ be the largest $y$ in interval $(0,{\epsilon_0}/{1+\epsilon_0}]$ which satisfies constraint (\ref{PCh1_DPCs}b). Then (\ref{PCh1_DPCs3}) is feasible if and only if $\tilde{y}_B$ exist and $F_{10}(\tilde{y}_B)\leq P_{max}-P_{11}(\epsilon_0/(1+\epsilon_0))$ ((\ref{PCh1_DPCs3})(c)).

Obviously, $\tilde{y}_B$ exists if and only if $P_{22}(0)<P_{max}$.
If problem (\ref{PCh1_DPCs3}) is feasible, then $\tilde{y}_B$ is the optimal $y$, i.e., $y_{opt}=\tilde{y}_B$.
\section{Proof for Proposition $1$}
We use proof by contradiction to prove Proposition $1$. Assume that the optimal solution belongs to the third case.

Since problem (\ref{Ch1_DPC1}) is feasible when $P_{20}>0$, from Corollary $2$,
it can be obtained that $P_{22}(\epsilon_0/(1+\epsilon_0))<P_{max})$.
Thus, from Theorem $3$, it can be obtained that:
\begin{align}
P_{10,opt}=F_{10}(y_{opt}),\quad P_{20,opt}=0, \quad x_{opt}=\frac{\epsilon_0}{1+\epsilon_0},\quad y_{opt}=\tilde{y}_B=\frac{\epsilon_0}{1+\epsilon_0}.
\end{align}

Let
 \Equ{\begin{cases}
P'_{10,opt}=\max\bigg\{0,F_{20}\left( \frac{\epsilon _0}{1+\epsilon _0} \right) ||\mathbf{h}_{20}||^2/||\mathbf{h}_{10}||^2\\
\quad\quad\quad\quad\quad\quad\quad-||\mathbf{h}_{20}||^2/||\mathbf{h}_{10}||^2\left( P_{max}-P_{22}(\epsilon_0/(1+\epsilon_0)) \right)\bigg\},
\\
P'_{20,opt}=P_{max}-P_{22}(\epsilon_0/(1+\epsilon_0)),\\
x'_{opt}=x_{opt},\quad y'_{opt}=y_{opt}.
\end{cases}}

It can be easily proved that the tuple $(P'_{10,opt}, P'_{20,opt}, x'_{opt}, y'_{opt})$ is a feasible solution of problem (\ref{Ch1_DPC1}), which belongs to Case I or II,
and is better than $(P_{10,opt}, P_{20,opt}, x_{opt}, y_{opt})$. This contradicts with the assumption that  $(P_{10,opt}, P_{20,opt}, x_{opt}, y_{opt})$ is the optimal solution and the proof is complete.
\section{Proof for Proposition 2}
The proof for Theorem (\ref{Pro2}) can be divided into three cases:

$1$) When {$x_{opt}=\tilde{x}_B$}:
For this case, we have $\tilde{x}_B<\frac{\epsilon_0}{\epsilon_0+||\mathbf{h}_{20}||^2/||\mathbf{h}_{10}||^2}$.
Assume that $P_{10,opt}$, $P_{20,opt}$ and $x_{opt}$ is not the optimal solution of problem (\ref{Ch1_DPC1}). Thus, there must exist a optimal solution $P'_{10}$, $P'_{20}$ and $x’$ such that:
\begin{align}{\label{A_asume1}}
    &P'_{10}+P'_{20}+P_{11}(x')<P_{20,opt}+P_{11}(x_{opt}),\\\notag
    &P'_{10}>0, P'_{20}>0.
\end{align}

Note that, $x'$ must be smaller than $\tilde{x}_B$, i.e., $x_{opt}$. Besides,
$P'_{10}$ and $P'_{20}$ should satisfy:
\Equ{\label{PCh1_sumpower}
    ||\mathbf{h}_{10}||^2P'_{10}+||\mathbf{h}_20||^2P'_{20}=||\mathbf{h}_{20}||^2F_{20}(x'),
}

According to $||\mathbf{h}_{20}||>||\mathbf{h}_{10}||$, we have
$P'_{10}+P'_{20}>F_{20}(x'),$
thus
\Equ{
   &\left(P'_{10}+P'_{20}+P_{11}(x')\right)-(P_{20,opt}+P_{11}(x_{opt}))\\\notag
   &>\left(F_{20}(x')+P_{11}(x')\right)-(P_{20,opt}+P_{11}(x_{opt}))>0
}
which is contradict with (\ref{A_asume1}).

$2$) {$x_{opt}=\tilde{x}_A$}:
Assume that $P_{10,opt}$, $P_{20,opt}$ and $x_{opt}$ is not the optimal solution of problem (\ref{Ch1_DPC1}). Thus, there must exist an optimal solution $P'_{10}$, $P'_{20}$ and $x’$ such that:
\begin{align}{\label{A_asume2}}
    P'_{10}+P'_{20}+P_{11}(x')<P_{20,opt}+P_{11}(x_{opt}).
\end{align}

Note that, similar to the proof for the case when $x_{opt}=\tilde{x}_B$, it can be proved that $x'$ must satisfy $x'<\tilde{x}_A$.
In that case, (\ref{PCh1_sumpower}) also needs to be satisfied. According to the fact that $F_{20}(x)$ is monotonically decreasing with $x$, and $F_{20}(\tilde{x}_A)=P_{max}-P_{22}(\epsilon_0/(1+\epsilon_0))$, the optimal $P_{20}$ should be $P'_{20}=P_{max}-P_{22}(\epsilon_0/(1+\epsilon_0))$, thus the following relationship can be obtained:
\Equ{
    &\quad\left(P'_{20}+P'_{10}+P_{11}(x')\right)-\left(P_{20,opt}+P_{11}(x_{opt})\right)\\\notag
    &=(P_{max}-P_{22}(\frac{\epsilon_0}{1+\epsilon_0})+P'_{10}+P_{11}(x'))-(P_{max}-P_{22}(\frac{\epsilon_0}{1+\epsilon_0})+P_{11}(x_{opt})\\\notag
    &=\left(\frac{||\mathbf{h}_{20}||^2(P_{max}-P_{22}(\frac{\epsilon_0}{1+\epsilon_0}))}{||\mathbf{h}_{10}||^2}+P'_{10}+P_{11}(x')\right)
    -\left(\frac{||\mathbf{h}_{20}||^2(P_{max}-P_{22}(\frac{\epsilon_0}{1+\epsilon_0}))}{||\mathbf{h}_{10}||^2}+P_{11}(x_{opt})\right)\\\notag
    &=\left({F_{20}(x')}/{||\mathbf{h}_{10}||^2}+P_{11}(x')\right)-\left({F_{20}(x_{opt})}/{||\mathbf{h}_{10}||^2}+P_{11}(x_{opt})\right)
.}

Define a new function:
$
\tilde{G}(x)={||\mathbf{h}_{20}||^2}/{||\mathbf{h}_{10}||^2}F_{20}(x)+P_{11}(x),
$
by taking the derivatives of $\tilde{G}(x)$, it can be concluded that $\tilde{G}(x)$ is monotonically decreasing with $x$ when $0<x\le\epsilon_0/(1+\epsilon_0)$, thus
\Equ{
    &\quad\left(P'_{20}+P'_{10}+P_{11}(x')\right)-\left(P_{20,opt}+P_{11}(x_{opt})\right)>0,
}
which leads to the contradiction.
$3$) {{$x_{opt}=\frac{\epsilon_0}{\epsilon_0+||\mathbf{h}_{20}||^2/||\mathbf{h}_{10}||^2}$}}:
For this case, the optimal solution is the same as the case where there is no power constraint, as shown in Corollary 1. Thus, there is no feasible solutions which are more optimal.
\bibliographystyle{IEEEtran}
\bibliography{IEEEabrv,ref}

\begin{thebibliography}{10}
\providecommand{\url}[1]{#1}
\csname url@samestyle\endcsname
\providecommand{\newblock}{\relax}
\providecommand{\bibinfo}[2]{#2}
\providecommand{\BIBentrySTDinterwordspacing}{\spaceskip=0pt\relax}
\providecommand{\BIBentryALTinterwordstretchfactor}{4}
\providecommand{\BIBentryALTinterwordspacing}{\spaceskip=\fontdimen2\font plus
\BIBentryALTinterwordstretchfactor\fontdimen3\font minus
  \fontdimen4\font\relax}
\providecommand{\BIBforeignlanguage}[2]{{%
\expandafter\ifx\csname l@#1\endcsname\relax
\typeout{** WARNING: IEEEtran.bst: No hyphenation pattern has been}%
\typeout{** loaded for the language `#1'. Using the pattern for}%
\typeout{** the default language instead.}%
\else
\language=\csname l@#1\endcsname
\fi
#2}}
\providecommand{\BIBdecl}{\relax}
\BIBdecl

\bibitem{irmer2011coordinated}
R.~Irmer, H.~Droste, P.~Marsch, M.~Grieger, G.~Fettweis, S.~Brueck, H.-P.
  Mayer, L.~Thiele, and V.~Jungnickel, ``{Coordinated multipoint: Concepts,
  performance, and field trial results},'' \emph{{IEEE} Commun. Mag.}, vol.~49,
  no.~2, pp. 102--111, Feb. 2011.

\bibitem{kim2021energy}
Y.~Kim, J.~Jeong, S.~Ahn, J.~Kwak, and S.~Chong, ``{Energy and Delay Guaranteed
  Joint Beam and User Scheduling Policy in 5G CoMP Networks},'' \emph{IEEE
  Trans. Wireless Commun.}, vol.~21, no.~4, pp. 2742--2756, Apr. 2021.

\bibitem{ding2017NOMAsurvey}
Z.~Ding, X.~Lai, G.~K. Karagiannidis, R.~Schober, J.~Yuan, and V.~Bhargava,
  ``{A survey on non-orthogonal multiple access for 5G networks: research
  challenges and future trends},'' \emph{{IEEE} J. Sel. Areas Commun.},
  vol.~35, no.~10, pp. 2181--2195, Oct. 2017.

\bibitem{choi2014non}
J.~Choi, ``{Non-orthogonal multiple access in downlink coordinated two-point
  systems},'' \emph{{IEEE} Commun. Lett.}, vol.~18, no.~2, pp. 313--316, Feb.
  2014.

\bibitem{sys2017nnomafeasibility}
Y.~Sun, Z.~Ding, X.~Dai, and G.~K. Karagiannidis, ``{A feasibility study on
  network NOMA},'' \emph{{IEEE} Trans. Commun.}, vol.~66, no.~9, pp.
  4303--4317, Sep. 2018.

\bibitem{new2020network}
W.~K. New, C.~Y. Leow, K.~Navaie, and Z.~Ding, ``{Network NOMA for Co-existence
  of Aerial and Terrestrial Users},'' in \emph{IEEE VTC2020-Fall}, Victoria,
  BC, Canada, Nov. 2020, pp. 1--5.

\bibitem{ali2018coordinated}
M.~S. Ali, E.~Hossain, and D.~I. Kim, ``{Coordinated multipoint transmission in
  downlink multi-cell NOMA systems: Models and spectral efficiency
  performance},'' \emph{IEEE Wireless Commun.}, vol.~25, no.~2, pp. 24--31,
  Apr. 2018.

\bibitem{makki2020survey}
B.~Makki, K.~Chitti, A.~Behravan, and M.-S. Alouini, ``{A survey of NOMA:
  Current status and open research challenges},'' \emph{IEEE Open J. Commun.
  Soc.}, vol.~1, pp. 179--189, Feb. 2020.

\bibitem{tian2016performance}
Y.~Tian, A.~R. Nix, and M.~Beach, ``{On the Performance of Opportunistic NOMA
  in Downlink CoMP Networks},'' \emph{{IEEE} Commun. Lett.}, vol.~20, no.~5,
  pp. 998--1001, May 2016.

\bibitem{al2019generalized}
Y.~Al-Eryani, E.~Hossain, and D.~I. Kim, ``{Generalized coordinated multipoint
  (GCoMP)-enabled NOMA: Outage, capacity, and power allocation},'' \emph{IEEE
  Trans. Commun.}, vol.~67, no.~11, pp. 7923--7936, Jul. 2019.

\bibitem{sys2019PCP}
Y.~Sun, Z.~Ding, X.~Dai, and O.~A. {Dobre}, ``{On the performance of network
  NOMA in uplink CoMP systems: a stochastic geometry approach},'' \emph{{IEEE}
  Trans. Commun.}, vol.~67, no.~7, pp. 5084--5098, Jul. 2019.

\bibitem{zhang2020performance}
Y.~Zhang, J.~Mu, and J.~Xiaojun, ``Performance of multi-cell mmwave noma
  networks with base station cooperation,'' \emph{IEEE Commun. Lett.}, vol.~25,
  no.~2, pp. 442--445, feb 2021.

\bibitem{elhattab2020comp}
M.~Elhattab, M.-A. Arfaoui, and C.~Assi, ``{CoMP transmission in downlink
  NOMA-based heterogeneous cloud radio access networks},'' \emph{IEEE Trans.
  Commun.}, vol.~68, no.~12, pp. 7779--7794, Dec. 2020.

\bibitem{sun2021outage}
Y.~Sun, Z.~Ding, and X.~Dai, ``{On the outage performance of network NOMA
  (N-NOMA) modeled by poisson line cox point process},'' \emph{{IEEE} Trans.
  Veh. Technol.}, vol.~70, no.~8, pp. 7936--7950, 8 2021.

\bibitem{liu2017power}
Z.~Liu, G.~Kang, L.~Lei, N.~Zhang, and S.~Zhang, ``{Power allocation for energy
  efficiency maximization in downlink CoMP systems with NOMA},'' in \emph{Proc.
  {IEEE} WCNC)}, {San Francisco, CA, USA}, 2017, pp. 1--6.

\bibitem{ali2018downlink}
M.~S. Ali, E.~Hossain, A.~Al-Dweik, and D.~I. Kim, ``{Downlink power allocation
  for CoMP-NOMA in multi-cell networks},'' \emph{IEEE Trans. Commun.}, vol.~66,
  no.~9, pp. 3982--3998, Sep. 2018.

\bibitem{elhattab2020power}
M.~Elhattab, M.~A. Arfaoui, and C.~Assi, ``{Power allocation in CoMP-empowered
  C-NOMA networks},'' \emph{IEEE Networking Lett.}, vol.~3, no.~1, pp. 10--14,
  Jan. 2020.

\bibitem{rezvani2021resource}
S.~Rezvani, N.~M. Yamchi, M.~R. Javan, and E.~A. Jorswieck, ``{Resource
  allocation in virtualized CoMP-NOMA HetNets: Multi-connectivity for joint
  transmission},'' \emph{IEEE Trans. Commun.}, vol.~69, no.~6, pp. 4172--4185,
  Jun. 2021.

\bibitem{elhattab2022joint}
M.~Elhattab, M.~A. Arfaoui, and C.~Assi, ``{Joint Clustering and Power
  Allocation in Coordinated Multipoint Assisted C-NOMA Cellular Networks},''
  \emph{IEEE Trans. Commun.}, 2022, to be published.

\bibitem{hedayati2020comp}
M.~Hedayati and I.-M. Kim, ``{CoMP-NOMA in the SWIPT networks},'' \emph{IEEE
  Trans. Wireless Commun.}, vol.~19, no.~7, pp. 4549--4562, Jul. 2020.

\bibitem{lei2020outage}
R.~Lei and D.~Xu, ``{On the outage performance of JT-CoMP-CNOMA networks with
  SWIPT},'' \emph{IEEE Commun. Lett.}, vol.~25, no.~2, pp. 432--436, Feb. 2020.

\bibitem{elhattab2020joint}
M.~Elhattab, M.~A. Arfaoui, and C.~Assi, ``{A Joint CoMP C-NOMA for Enhanced
  Cellular System Performance},'' \emph{IEEE Commun. Lett.}, vol.~24, no.~9,
  pp. 1919--1923, Sep. 2020.

\bibitem{elhattab2022ris}
M.~Elhattab, M.~A. Arfaoui, C.~Assi, and A.~Ghrayeb, ``{RIS-Assisted Joint
  Transmission in a Two-Cell Downlink NOMA Cellular System},'' \emph{{IEEE} J.
  Select. Areas Commun.}, vol.~40, no.~4, pp. 1270--1286, apr 2022.

\bibitem{wang2020power}
H.~Wang, C.~Liu, Z.~Shi, Y.~Fu, and R.~Song, ``{Power minimization for two-cell
  IRS-aided NOMA systems with joint detection},'' \emph{IEEE Commun. Lett.},
  vol.~25, no.~5, pp. 1635--1639, 5 2020.

\bibitem{hou2021joint}
T.~Hou, J.~Wang, Y.~Liu, X.~Sun, A.~Li, and B.~Ai, ``{A Joint Design for
  STAR-RIS Enhanced NOMA-CoMP Networks: A
  Simultaneous-Signal-Enhancement-and-Cancellation-Based (SSECB) Design},''
  \emph{{IEEE} Trans. Veh. Technol.}, vol.~71, no.~1, pp. 1043--1048, Jan.
  2021.

\bibitem{shin2016coordinated}
W.~Shin, M.~Vaezi, B.~Lee, D.~J. Love, J.~Lee, and H.~V. Poor, ``{Coordinated
  beamforming for multi-cell MIMO-NOMA},'' \emph{IEEE Commun. Lett.}, vol.~21,
  no.~1, pp. 84--87, Jan. 2016.

\bibitem{nguyen2017precoder}
V.-D. Nguyen, H.~D. Tuan, T.~Q. Duong, H.~V. Poor, and O.-S. Shin, ``{Precoder
  design for signal superposition in MIMO-NOMA multicell networks},''
  \emph{IEEE J. Selected Areas Commun}, vol.~35, no.~12, pp. 2681--2695, Dec.
  2017.

\bibitem{sun2018joint}
X.~Sun, N.~Yang, S.~Yan, Z.~Ding, D.~W.~K. Ng, C.~Shen, and Z.~Zhong, ``{Joint
  beamforming and power allocation in downlink NOMA multiuser MIMO networks},''
  \emph{IEEE Trans. Wireless Commun.}, vol.~17, no.~8, pp. 5367--5381, Aug.
  2018.

\bibitem{fu2020zero}
Y.~Fu, M.~Zhang, L.~Sala{\"u}n, C.~W. Sung, and C.~S. Chen, ``{Zero-forcing
  oriented power minimization for multi-cell MISO-NOMA systems: A joint user
  grouping, beamforming, and power control perspective},'' \emph{IEEE J.
  Selected Areas Commun.}, vol.~38, no.~8, pp. 1925--1940, Aug. 2020.

\bibitem{chen2016application}
Z.~Chen, Z.~Ding, X.~Dai, and G.~K. Karagiannidis, ``{On the application of
  quasi-degradation to MISO-NOMA downlink},'' \emph{IEEE Trans. Signal
  Process.}, vol.~64, no.~23, pp. 6174--6189, Dec. 2016.

\bibitem{zhu2020optimal}
J.~Zhu, J.~Wang, Y.~Huang, K.~Navaie, Z.~Ding, and L.~Yang, ``{On optimal
  beamforming design for downlink MISO NOMA systems},'' \emph{IEEE Trans. Veh.
  Technol.}, vol.~69, no.~3, pp. 3008--3020, Mar. 2020.

\bibitem{liu2021quasi}
K.-H. Liu, ``{Quasi-Degradation Probability of Two-User NOMA Over Rician Fading
  Channels},'' \emph{IEEE Trans. Veh. Technol.}, vol.~70, no.~4, pp.
  3514--3524, Apr. 2021.

\end{thebibliography}
\end{document}